\RequirePackage[displaymath]{lineno}
\documentclass[aps,prd,twocolumn,showpacs,amsmath,amssymb]{revtex4-1}
\usepackage{epsfig}
\usepackage{graphicx}
\usepackage{dcolumn}
\usepackage{bm}
\usepackage{overpic}
\usepackage{subfigure}
\usepackage{float}
\usepackage{color}
\usepackage{amsmath}
\usepackage{mathcomp}
\usepackage{mathrsfs}
\usepackage{multirow}
\usepackage{rotating}

\begin{document}
\normalsize
\parskip=5pt plus 1pt minus 1pt

\title{ \boldmath Amplitude analysis of $D^{0} \rightarrow K^{-} \pi^{+} \pi^{+} \pi^{-}$ }
\vspace{-1cm}

\author{
   \begin{small}
    \begin{center}
      M.~Ablikim$^{1}$, M.~N.~Achasov$^{9,e}$, S. ~Ahmed$^{14}$, X.~C.~Ai$^{1}$, O.~Albayrak$^{5}$, M.~Albrecht$^{4}$, D.~J.~Ambrose$^{44}$, A.~Amoroso$^{49A,49C}$, F.~F.~An$^{1}$, Q.~An$^{46,a}$
% ...
         \\
         \vspace{0.2cm}
   (BESIII Collaboration)\\
      \vspace{0.2cm} {\it
         $^{1}$ Institute of High Energy Physics, Beijing 100049, People's Republic of China\\
% ...
$^{4}$ Bochum Ruhr-University, D-44780 Bochum, Germany\\
         $^{5}$ Carnegie Mellon University, Pittsburgh, Pennsylvania 15213, USA\\
% ...
$^{9}$ G.I. Budker Institute of Nuclear Physics SB RAS (BINP), Novosibirsk 630090, Russia\\
% ...
$^{14}$ Helmholtz Institute Mainz, Johann-Joachim-Becher-Weg 45, D-55099 Mainz, Germany\\
% ...
$^{44}$ University of Rochester, Rochester, New York 14627, USA\\
% ...
$^{46}$ University of Science and Technology of China, Hefei 230026, People's Republic of China\\
% ...
$^{49}$ (A)University of Turin, I-10125, Turin, Italy; (B)University of Eastern Piedmont, I-15121, Alessandria, Italy; (C)INFN, I-10125, Turin, Italy\\
% ...
$^{a}$ Also at State Key Laboratory of Particle Detection and Electronics, Beijing 100049, Hefei 230026, People's Republic of China\\
% ...
$^{e}$ Also at the Novosibirsk State University, Novosibirsk, 630090, Russia\\
% ...
         }\end{center}
      \vspace{0.4cm}
      \end{small}
}

\affiliation{}
\vspace{-4cm}
\begin{abstract}
We present an amplitude analysis of the decay $D^{0} \rightarrow K^{-} \pi^{+} \pi^{+} \pi^{-}$ based on 
a data sample of 2.93 ${\mbox{\,fb}^{-1}}$ acquired by the BESIII detector at the $\psi(3770)$ resonance.
With a nearly background free sample of about 16000 events, 
we investigate the substructure of the decay and  
determine the relative fractions and the phases among the different intermediate processes. 
Our amplitude model includes the two-body decays $D^{0} \rightarrow \bar{K}^{*0}\rho^{0}$, 
$D^{0} \rightarrow K^{-}a_{1}^{+}(1260)$ and $D^{0} \rightarrow K_{1}^{-}(1270)\pi^{+}$, 
the three-body decays $D^{0} \rightarrow \bar{K}^{*0}\pi^{+}\pi^{-}$ and
$D^{0} \rightarrow K^{-}\pi^{+}\rho^{0}$, as well as 
the four-body nonresonant decay $D^{0} \rightarrow K^{-}\pi^{+}\pi^{+}\pi^{-}$.
The dominant intermediate process is $D^{0} \rightarrow K^{-}a_{1}^{+}(1260)$, 
accounting for a fit fraction of $54.6\%$. 
\end{abstract}
\maketitle

\section{Introduction}
\label{sec:introduction}
The decay $D^{0} \rightarrow K^{-} \pi^{+} \pi^{+} \pi^{-}$ is one of the three
golden decay modes of the neutral $D$ meson 
(the other two are $D^{0} \rightarrow K^{-} \pi^{+}$ 
and $D^{0} \rightarrow K^{-} \pi^{+} \pi^{0}$).
Due to a large branching fraction and low background it is well suited 
to use as a reference channel for other decays of the $D^{0}$ meson~\cite{PDG}.
An accurate knowledge of its resonant substructure and the relative amplitudes and phases are important to
reduce systematic uncertainties in analyses that use this channel for reference.
In particular, the lack of knowledge of the substructure leads to one of the largest systematic uncertainties
in the measurement of the absolute branching fractions of the $D$ hadronic decays~\cite{CLEODdecay}.
The knowledge of the decay substructure in combination with a precise measurement
of strong phases can also help to improve the measurement of the CKM angle 
$\gamma$ (the phase of $V_{cb}$ relative to $V_{ub}$)~\cite{ADS}.
In the measurement of $\gamma$, the parametrization model is an important input information in a model dependent
method and also can be used to generate Monte Carlo (MC) simulations to check the sensitivity
in a model independent method~\cite{K3Pigamma}.
Furthermore, the branching fractions of intermediate processes can be used to
understand the $D^{0}-\bar{D}^{0}$ mixing in theory~\cite{AFFalk,HYCheng}.

The decay $D^{0} \rightarrow K^{-} \pi^{+} \pi^{+} \pi^{-}$ was studied by
Mark III~\cite{MarKIII} and E691~\cite{E691} more than twenty years ago.
Both measurements are affected by low statistics.
Using about $1300$ signal events,
Mark III obtained the branching fractions for $D^{0} \rightarrow K^{-} a^{+}_{1}(1260)$,
$D^{0} \rightarrow \bar{K}^{*0} \rho^{0}$, $D^{0} \rightarrow K_{1}^{-}(1270)\pi^{+}$,
as well as for the three- and four-body nonresonant decays.
Based on $1745$ signal events and $800$ background events,
E691 obtained a similar result but without considering the $D^{0} \rightarrow K^{-}_{1}(1270)\pi^{+}$ decay mode.
The results from Mark~III and E691 have large uncertainties.
Therefore, further experimental study of $D^{0} \rightarrow K^{-} \pi^{+} \pi^{+} \pi^{-}$
decay is of great importance for improving the precision of future measurements.

In this paper, a data sample of about 2.93 ${\mbox{\,fb}^{-1}}$~\cite{datasample, datasample2}
collected at the $\psi(3770)$ resonance with the BESIII detector in 2010 and 2011 is used.
We perform an amplitude analysis of the decay $D^{0} \rightarrow K^{-} \pi^{+} \pi^{+} \pi^{-}$
(the inclusion of charge conjugate reactions is implied) to study the
resonant substructure in this decay.
The $\psi(3770)$ decays into a $D^{0}\bar{D}^{0}$ pair without any
additional hadrons.
We employ a double-tag method to measure the branching fraction.
In order to suppress the backgrounds from other charmed meson decays and 
continuum (QED and $q\bar{q}$) processes, 
only the decay mode $\bar{D}^{0} \rightarrow K^{+}\pi^{-}$
is used to tag the $D^{0}\bar{D}^{0}$ pair.
A detailed discussion of background can be found in Sec.~\ref{sec:evt}.
The amplitude model is constructed using the covariant tensor formalism~\cite{Zou}.

\section{Detection and Data Sets}
\label{sec:Data Set and Event Seletion}

The BESIII detector is described in detail in Ref.~\cite{detector}. The geometrical acceptance
of the BESIII detector is 93\% of the full solid angle.
Starting from the interaction point (IP), it consists of a main drift chamber (MDC),
a time-of-flight (TOF) system, a CsI(Tl) electromagnetic calorimeter (EMC) and a muon system (MUC) with
layers of resistive plate chambers (RPC) in the iron return yoke of a 1.0 T superconducting solenoid.
The momentum resolution for charged tracks in the MDC is 0.5\% at a transverse momentum of 1 GeV$/c$.

Monte Carlo (MC) simulations are based on GEANT4~\cite{sim}.
The production of $\psi(3770)$ is simulated with the KKMC~\cite{KKMC}
package, taking into account the beam energy spread and initial-state
radiation (ISR).
The PHOTOS~\cite{FSR} package is used to simulate the final-state radiation (FSR) of charged tracks.
The MC samples, which consist of $\psi(3770)$ decays to $D\bar{D}$, non-$D\bar{D}$,
ISR production of low mass charmonium states and continuum processes,
are referred to as ``generic MC" samples.
The EvtGen~\cite{EvtGen} package is used to simulate
the known decay modes with branching fractions taken from
the Particle Data Group (PDG)~\cite{PDG}, and the remaining
unknown decays are generated with the LundCharm model~\cite{LundCharm}.
The effective luminosities of the generic MC samples correspond to at least 5 times
the data sample luminosity. They are used to investigate possible backgrounds.
The decay $D^{0} \rightarrow K_{S}^{0}(\pi^{+}\pi^{-})K^{-}\pi^{+}$ has
the same final state as signal and is investigated using a dedicated MC sample with
the decay chain of $\psi(3770)\rightarrow D^{0}\bar{D}^{0}$ with
$D^{0} \rightarrow K_{S}^{0}K^{-}\pi^{+}$ and $\bar{D}^{0} \rightarrow K^{+}\pi^{-}$,
referred to as the ``$K_{S}^{0}K\pi$ MC". The decay model of $D^{0} \rightarrow K_{S}^{0}K^{-}\pi^{+}$ is
generated according to CLEO's results~\cite{KsKPi}.
In amplitude analysis, two sets of signal MC samples using different decay models are generated.
One sample is generated with an uniform distribution in phase space
for the $D^{0} \rightarrow K^{-}\pi^{+}\pi^{+}\pi^{-}$ decay, which is used to calculate the MC 
integrations and called the ``PHSP MC" sample.
The other sample is generated according to the results obtained in this analysis for
the $D^{0} \rightarrow K^{-}\pi^{+}\pi^{+}\pi^{-}$ decay. It is used to
check the fit performance, calculate the goodness of fit and estimate the detector efficiency,
and is called the ``SIGNAL MC" sample.

\section{Event Selection}
\label{sec:evt}
Good charged tracks are required to have a point of closest approach
to the interaction point (IP) within $10$ cm along the beam axis and
within $1$ cm in the plane perpendicular to beam.
The polar angle $\theta$ between the track and the $e^+$ beam direction
is required to satisfy $|\cos \theta|<0.93$.
Charged particle identification (PID) is implemented by combining the
energy loss ($dE/dx$) in the MDC and the time-of-fight information from the TOF. Probabilities
$P(K)$ and $P(\pi)$ with the hypotheses of $K$ or $\pi$ are then calculated.
Tracks without PID information are rejected.
Charged kaon candidates are required to have $P(K) > P(\pi)$,
while the $\pi$ candidates are required to have $P(\pi) > P(K)$.
The average efficiencies for the kaon and pions in $K^{-}\pi^{+}\pi^{+}\pi^{-}$ are 
$\sim98$\% and $\sim99$\% respectively.
The $D^{0}\bar{D^{0}}$ pair with $\bar{D}^{0} \rightarrow K^{+}\pi^{-}$ and
$D^{0} \rightarrow K^{-}\pi^{+}\pi^{+}\pi^{-}$ is reconstructed with the requirement
that the two $D^{0}$ mesons have opposite charm and do not have any tracks in common. 
Since the tracks in $K^{-}\pi^{+}\pi^{+}\pi^{-}$ have distinct momenta from those 
in $K^{+}\pi^{-}$,  misreconstructed signal events and $K/\pi$ particle misidentification are negligible.
Furthermore, a vertex fit with the hypothesis that all tracks originate from the IP
is performed, and the $\chi^{2}$ of the fit is required to be less than $200$.

\begin{figure*}[hbtp]
\begin{center}
\centering
\begin{minipage}[b]{0.42\textwidth}
\epsfig{width=0.70\textwidth,clip=true,file=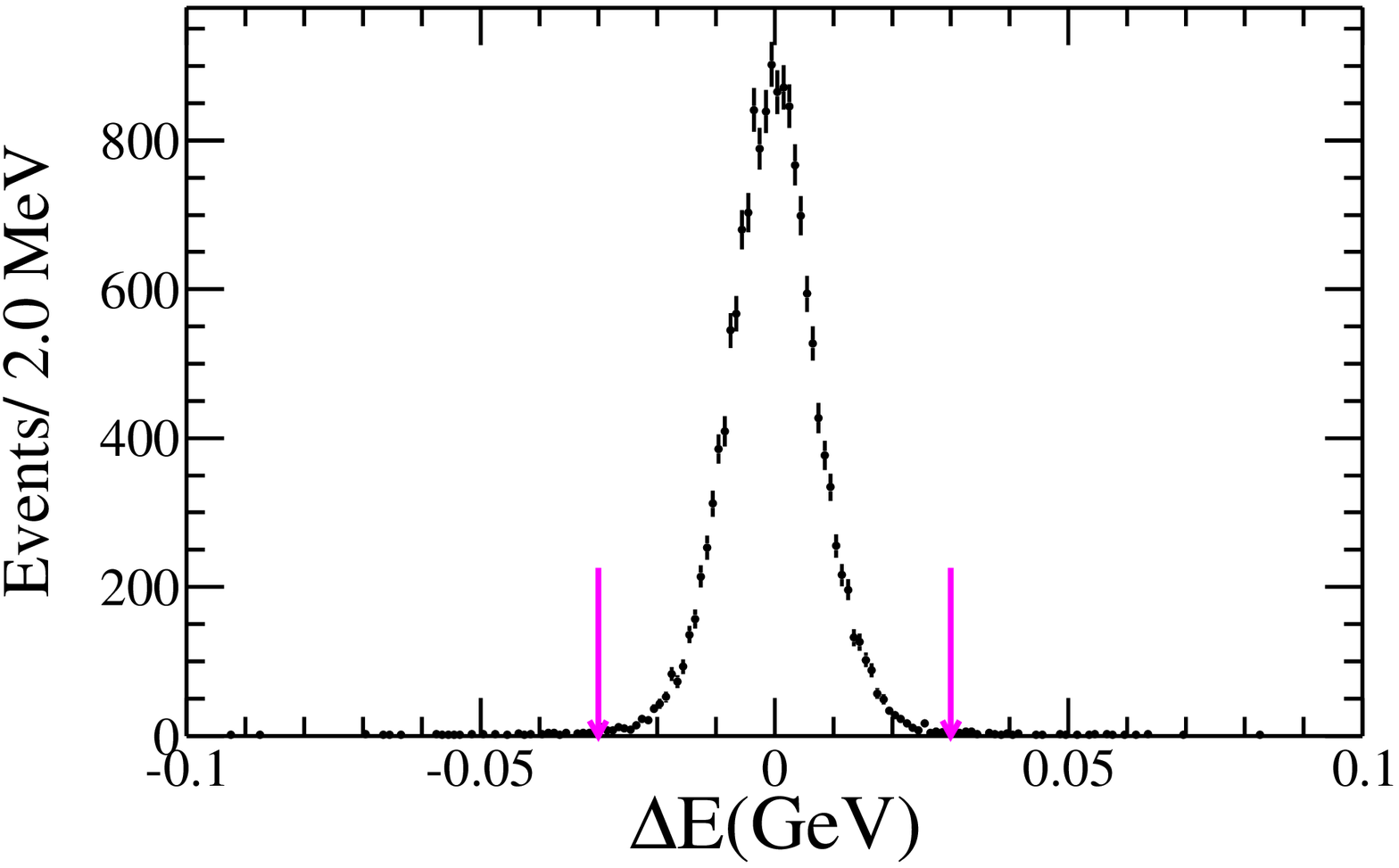}
\put(-120,75){(a)}
\end{minipage}
\begin{minipage}[b]{0.42\textwidth}
\epsfig{width=0.70\textwidth,clip=true,file=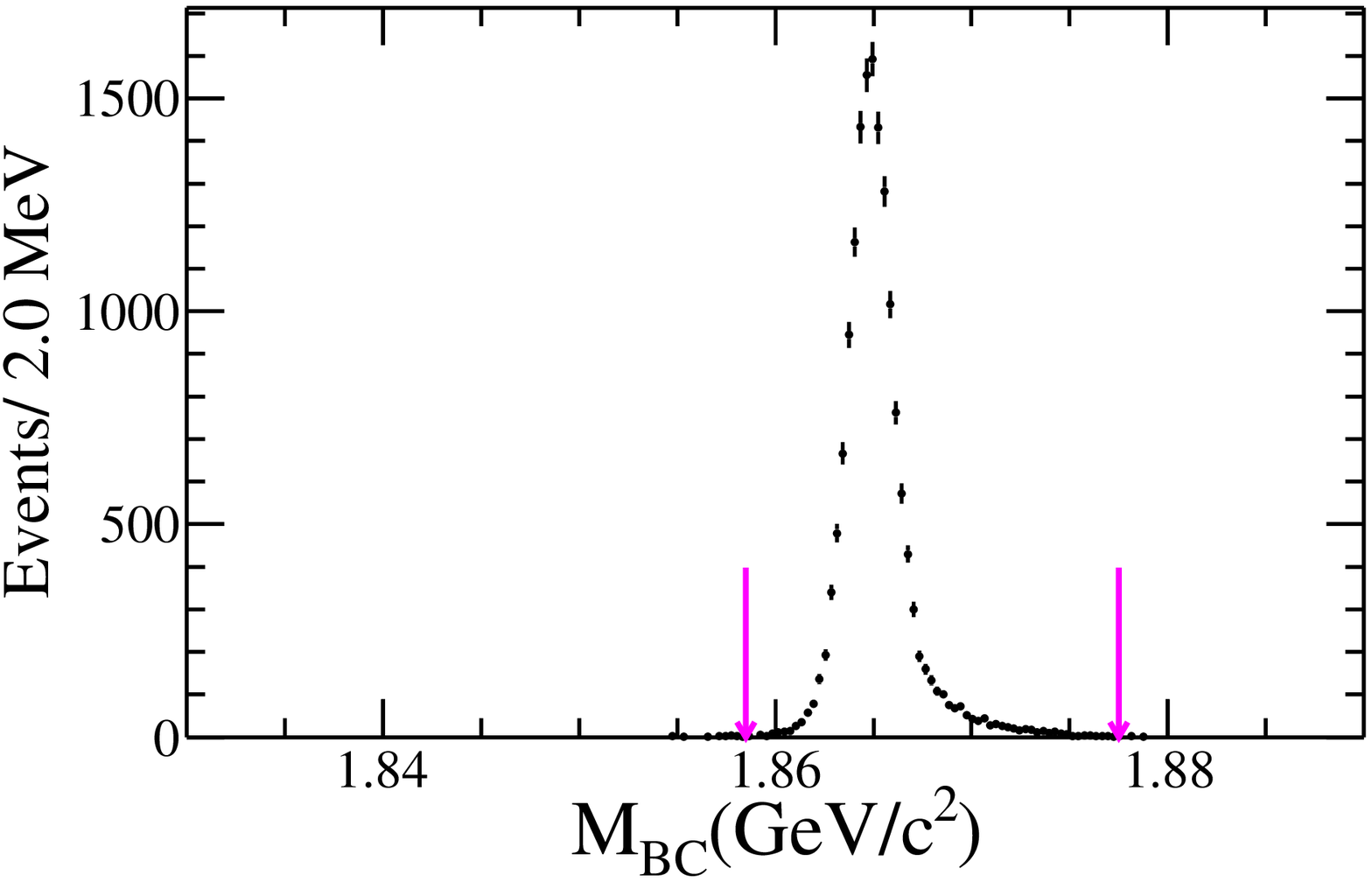}
\put(-120,75){(b)}
\end{minipage}
\begin{minipage}[b]{0.42\textwidth}
\epsfig{width=0.70\textwidth,clip=true,file=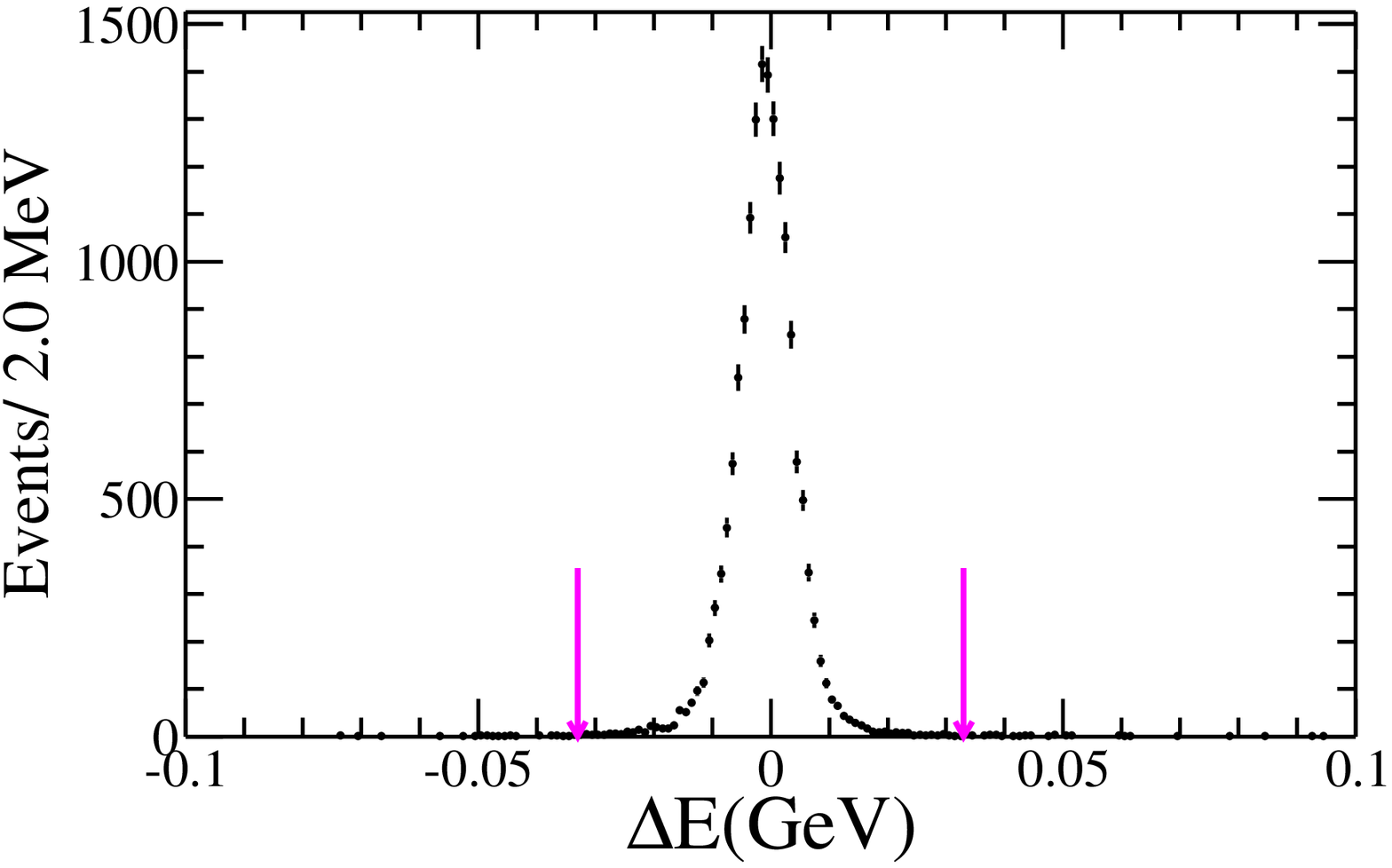}
\put(-120,75){(c)}
\end{minipage}
\begin{minipage}[b]{0.42\textwidth}
\epsfig{width=0.70\textwidth,clip=true,file=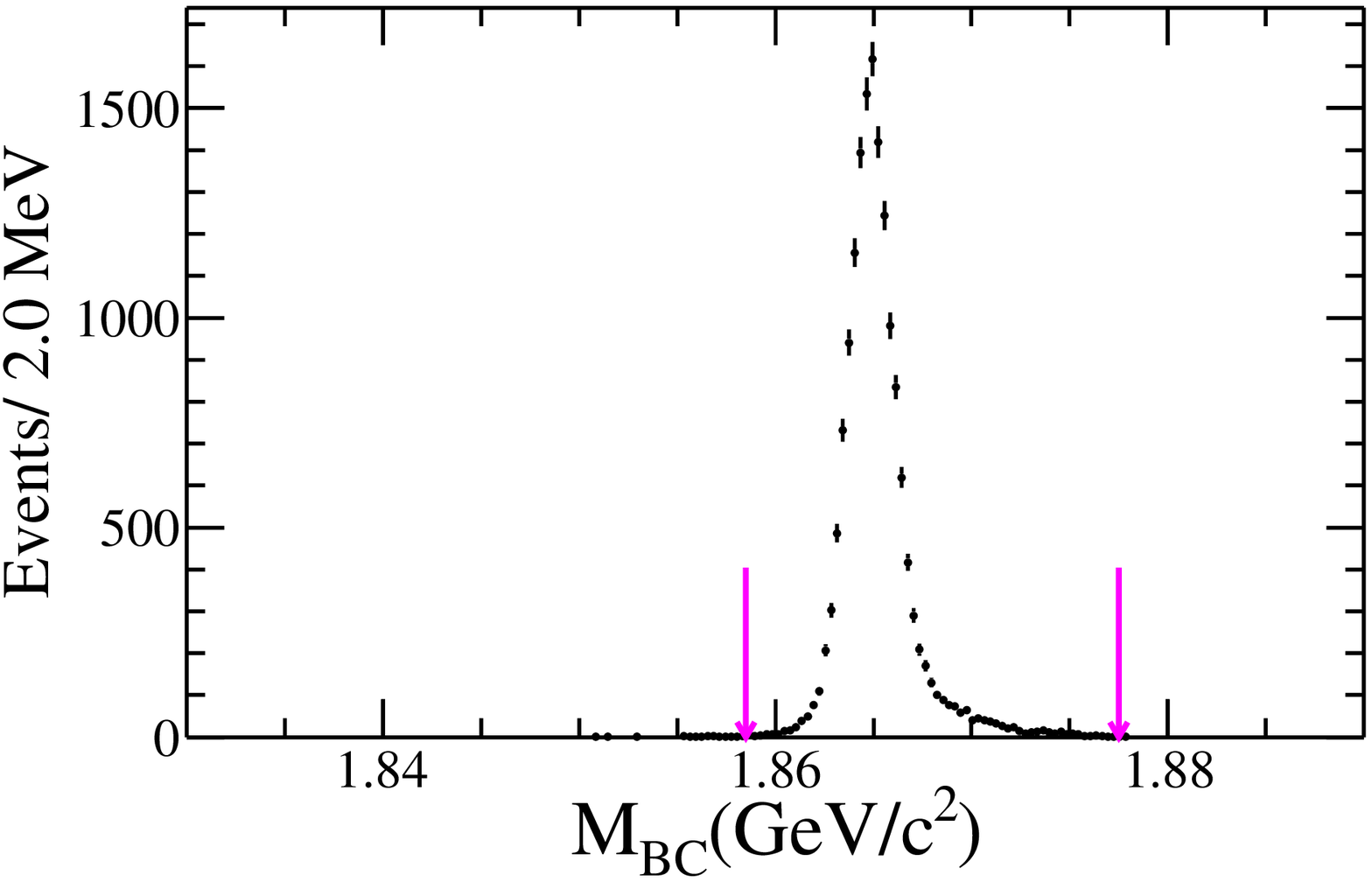}
\put(-120,75){(d)}
\end{minipage}
\caption{Distributions of data for $\Delta E$ [(a) and (c)] and $M_{{\rm BC}}$
[(b) and (d)] in $K^{+}\pi^{-}$ side [(a) and (b)]
and in $K^{-}\pi^{+}\pi^{+}\pi^{-}$ side [(c) and (d)]. The arrows indicate the selection criteria.
In each plot, all selection criteria described in this section
have been applied except the one on the variable.}
\label{fig:selection}
\end{center}
\end{figure*}

For the $K^{+}\pi^{-}$ and $K^{-}\pi^{+}\pi^{+}\pi^{-}$ combinations,
two variables, $M_{{\rm BC}}$ and $\Delta E$, are calculated:
\begin{eqnarray}
\begin{aligned}
 M_{{\rm BC}} \equiv \sqrt{E_{{\rm beam}}^2 -  \vec p_{D}^2},\\
\end{aligned}
\end{eqnarray}
and
\begin{eqnarray}
\begin{aligned}
\Delta E \equiv E_D - E_{{\rm beam}},
\end{aligned}
\end{eqnarray}
where $\vec p_D$ and $E_D$ are the reconstructed momentum and energy of a $D$ candidate,
$E_{{\rm beam}}$ is the calibrated beam energy.
The signal events form a peak around zero in the $\Delta E$ distribution
 and around the $D^{0}$ mass in the $M_{{\rm BC}}$ distribution.
We require $-0.03 < \Delta E < 0.03$~GeV for the $K^{+}\pi^{-}$ final state,
$-0.033 < \Delta E < 0.033$~GeV for the $K^{-}\pi^{+}\pi^{+}\pi^{-}$ final state
and $1.8575 < M_{{\rm BC}} < 1.8775$~GeV/$c^{2}$
for both of them. The corresponding $\Delta E$ and $M_{{\rm BC}}$ of selected candidate
are shown in Fig.~\ref{fig:selection}, where the background is negligible.

To ensure the $D^{0}$ meson is on shell and improve the resolution, 
the selected candidate events are further
subjected to a five-constraint (5C) kinematic fit, which constrains the
total four-momentum of all final state particles to the initial four-momentum of the $e^{+}e^{-}$ system,
and the invariant mass of signal side $K^{-}\pi^{+}\pi^{+}\pi^{-}$
constrains to the $D^{0}$ mass in PDG~\cite{PDG}.
We discard events with a $\chi^{2}$ of the 5C kinematic fit larger than $40$.
In order to suppress the background of $D^{0} \rightarrow K_{S}^{0}K^{-}\pi^{+}$ with
$K_{S}^{0} \rightarrow \pi^{+}\pi^{-}$, which has the same final state
as our signal decay, we perform a vertex constrained fit on any
$\pi^{+}\pi^{-}$ pair in the signal side if the $\pi^{+}\pi^{-}$ invariant mass falls into the mass window
$|m_{\pi^{+}\pi^{-}}-m_{K_{S}^{0}}|<0.03$~GeV/$c^{2}$
($m_{K_{S}^{0}}$ is the $K_{S}^{0}$ nominal mass~\cite{PDG}),
and reject the event if the corresponding significance of decay length
($e.g.$ the distance of the decay vertex to IP) is larger than 2$\sigma$.
The $K_{S}^{0}$ veto eliminates about 80\% $D^{0} \rightarrow K_{S}^{0} K^{-}\pi^{+}$ background
while retaining about 99\% of signal events.
After applying all selection criteria, 15912 candidate events are obtained with a purity of 99.4$\%$,
as estimated by MC simulation.

The MC studies indicate that the dominant background arises from the
$D^{0} \rightarrow K_{S}^{0}K^{-}\pi^{+}$ decay, the corresponding
produced number of events is estimated according to
\begin{eqnarray}
\begin{aligned}
\label{Num_KsKPi}
&N(K_{S}^{0}K^{-}\pi^{+}|K^{+}\pi^{-})= \\
&\frac{Y(K^{-}\pi^{+}\pi^{+}\pi^{-}|K^{+}\pi^{-})}
{\epsilon(K^{-}\pi^{+}\pi^{+}\pi^{-}|K^{+}\pi^{-})}
\times \frac{\mathcal{B}(K_{S}^{0}K^{-}\pi^{+})}{\mathcal{B}(K^{-}\pi^{+}\pi^{+}\pi^{-})},
\end{aligned}
\end{eqnarray}
where $N(K_{S}^{0}K^{-}\pi^{+}|K^{+}\pi^{-})$ is the production
of $\psi(3770) \rightarrow D^{0}\bar{D}^{0}$
with $D^{0} \rightarrow K_{S}^{0}K^{-}\pi^{+}$ and $\bar{D}^{0} \rightarrow K^{+}\pi^{-}$,
$Y(K^{-}\pi^{+}\pi^{+}\pi^{-}|K^{+}\pi^{-})$ is the signal yield with background subtracted
but without efficiency correction applied and $\epsilon$ is the corresponding efficiency obtained
from the SIGNAL MC sample, which is generated according to the results of fit to data
whose peaking background estimated from the generic MC sample.
$\mathcal{B}(K^{-}\pi^{+}\pi^{+}\pi^{-})$ and $\mathcal{B}(K_{S}^{0}K^{-}\pi^{+})$
are the branching fractions for $D^{0} \rightarrow K^{-}\pi^{+}\pi^{+}\pi^{-}$ and
$D^{0} \rightarrow K_{S}^{0}K^{-}\pi^{+}$, respectively, which are quoted from the PDG~\cite{PDG}.
According to Eq.~(\ref{Num_KsKPi}), the number
of peaking background events ($N_\text{peaking}$) is estimated to be $96.8\pm14.5$.

All other backgrounds from $D\bar{D}$, $q\bar{q}$ and non-$D\bar{D}$
decays are studied with the generic MC sample.
Their total contribution is estimated to be less than ten events, of which 5.5 and 2.0 are
from the $D^{0}\bar{D}^{0}$ decays and the non-$D\bar{D}$ decays, respectively.
These backgrounds are neglected in the following analysis and their effect is considered as a systematic
uncertainty, as discussed in Sec.~\ref{BKG_sys}.

\section{Amplitude Analysis}
\label{Amplitude Analysis}
The decay modes which may contribute to the $D^{0} \rightarrow K^{-}\pi^{+}\pi^{+}\pi^{-}$ decay
are listed in Table~\ref{tab:spin},
where the symbols S, P, V, A, and T denote a scalar,
pseudoscalar, vector, axial-vector, and tensor state, respectively.
The letters $S$, $P$, and $D$ in square brackets refer to the relative
angular momentum between the daughter particles.
The amplitudes and the relative phases between the different decay modes
are determined with a maximum likelihood fit.
\subsection{Likelihood function construction}
\label{Likelihood construction}
The likelihood function is the product of the probability density function (PDF) of the observed events.
The signal PDF $f_{S}(p_{j})$ is given by
\begin{eqnarray}
\begin{aligned}
\label{signalPDF}
f_{S}(p_{j}) = \frac{\epsilon(p_{j})|M(p_{j})|^{2}R_{4}(p_{j})}
{\int \epsilon(p_{j})|M(p_{j})|^{2}R_{4}(p_{j})dp_{j}},
\end{aligned}
\end{eqnarray}
where $\epsilon(p_{j})$ is the detection efficiency parametrized in terms of the final four-momenta $p_{j}$.
The index $j$ refers to the different particles in the final state.
$R_{4}(p_{j})dp_{j}$ is the standard element of the four-body phase space~\cite{Zou}, which is given by
\begin{eqnarray}
\begin{aligned}
R_{4}(p_{j})dp_{j} =
\delta^{4}\left(p_{D^{0}}-\sum^{4}_{j=1}p_{j}\right)
\,\prod^{4}_{j=1}\frac{d^{3}{\bf p}_{j}}{(2\pi)^{3}2E_{j}}. 
\end{aligned}
\end{eqnarray}
$M(p_{j})$ is the total decay amplitude which is modeled as a
coherent sum over all contributing amplitudes
\begin{eqnarray}
\begin{aligned}
M(p_{j}) = \sum_{n}c_{n}A_{n}(p_{j}),
\end{aligned}
\end{eqnarray}
where the complex coefficient $c_{n}= \rho_{n}e^{i\phi_{n}}$ ($\rho_{n}$ and $\phi_{n}$
are the magnitude and phase for the $n^\text{th}$ amplitude, respectively)
and $A_{n}(p_{j})$ describe the relative contribution and the dynamics of the $n^\text{th}$ amplitude.
In four-body decays, the intermediate amplitude can be  
a quasi-two-body decay or a cascade decay amplitude, and $A_{n}(p_{j})$ is given by
\begin{eqnarray}
\begin{aligned}
A_{n}(p_{j}) = P_{n}^{1}(m_{1})P_{n}^{2}(m_{2})S_{n}(p_{j})F_{n}^{1}(p_{j})F_{n}^{2}(p_{j})F_{n}^{D}(p_{j}),
\end{aligned}
\end{eqnarray}
where the indices 1 and 2 correspond to the two intermediate resonances.
Here, $P_{n}^{\alpha}(m_{\alpha})$ and $F_{n}^{\alpha}(p_{j})$ ($\alpha = 1,2$)
are the propagator and the Blatt-Weisskopf barrier
factor~\cite{Blatt}, respectively, and
$F_{n}^{D}(p_{j})$ is the Blatt-Weisskopf barrier factor of the $D^{0}$
decay.
The parameters $m_{1}$ and $m_{2}$ in the propagators are the invariant masses of the corresponding systems.
For nonresonant states with orbital angular momentum between the daughters,
we set the propagator to unity, which can be regarded as a very broad resonance.
The spin factor $S_{n}(p_{j})$ is constructed with the covariant tensor formalism~\cite{Zou}.
In practice, the presence of the two $\pi^{+}$ mesons imposes a
Bose symmetry in the $K^{-}\pi^{+}\pi^{+}\pi^{-}$ final state.
This symmetry is explicitly accounted for in the amplitude by
exchange of the two pions with the same charge.

The contribution from the background is subtracted in the
likelihood calculation by assigning a negative weight to the background events
\begin{eqnarray}
\begin{aligned}
\ln L = \sum_{k=1}^{N_{{\rm data}}}\ln f_{S}(p_{j}^{k}) 
      - \sum_{k^{\prime}=1}^{N_{{\rm bkg}}}w_{k^{\prime}}^{{\rm bkg}}\ln f_{S}(p_{j}^{k^{\prime}}),
\end{aligned}
\end{eqnarray}
where $N_{{\rm data}}$ is the number of candidate events in data,
$w_{k^{\prime}}^{{\rm bkg}}$ and $N_{{\rm bkg}}$ are the weight and
the number of events from the background MC sample, respectively.
In the nominal fit, only the peaking background
$D^{0} \rightarrow K_{S}^{0}K^{-}\pi^{+}$ is considered, and the weight $w_{k^{\prime}}^{{\rm bkg}}$ is
fixed to ${N_\text{peaking}}/{N_\text{bkg}}$.
$p_{j}^{k}$ and $p_{j}^{k^{\prime}}$ are the four-momenta of the $j^{{\rm th}}$
final particle in the $k^{{\rm th}}$ event of the data sample and in the $k^{\prime {\rm th}}$
event of the background MC sample, respectively.

The normalization integral is determined by a MC technique taking into account the difference of detector
efficiencies for PID and tracking between data and MC simulation.
The weight for a given MC event is defined as
\begin{eqnarray}
\begin{aligned}
\label{trackingPID}
\gamma_{\epsilon}(p_{j}) = \prod_{j}{\frac{\epsilon_{j,{\rm data}}(p_{j})}{\epsilon_{j,{\rm MC}}(p_{j})}},
\end{aligned}
\end{eqnarray}
where $\epsilon_{j,{\rm data}}(p_{j})$ and $\epsilon_{j,{\rm MC}}(p_{j})$ are the PID or
tracking efficiencies for charged tracks as a function of $p_{j}$ for the data and MC sample, respectively.
The efficiencies $\epsilon_{j,{\rm data}}(p_{j})$ and $\epsilon_{j,{\rm MC}}(p_{j})$ are determined by studying the
$D^{0} \rightarrow K^{-}\pi^{+}\pi^{+}\pi^{-}$ sample for data and the MC sample respectively.
The MC integration is then given by
\begin{eqnarray}
\begin{aligned}
&\int \epsilon(p_{j})|M(p_{j})|^{2}R_{4}(p_{j})dp_{j}\\ 
&=\frac{1}{N_{{\rm MC}}}\sum_{k_{{\rm MC}}}^{N_{{\rm MC}}}
  \frac{|M(p_{j}^{k_{{\rm MC}}}))|^{2}\gamma_{\epsilon}(p_{j}^{k_{{\rm MC}}})}
  {|M^{{\rm gen}}(p_{j}^{k_{{\rm MC}}})|^{2}},
\end{aligned}
\end{eqnarray}
where $k_{{\rm MC}}$ is the index of the $k_{{\rm MC}}^{{\rm th}}$ event of the MC sample and
$N_{{\rm MC}}$ is the number of the selected MC events.
$M^{{\rm gen}}(p_{j})$ is the PDF function used to generate the MC samples in MC integration.
In the numerator of Eq.~(\ref{signalPDF}), $\epsilon(p_{j})$ is independent of the fitted variables,
so it is regarded as a constant term in the fit.

\subsubsection{Spin factors}
Due to the limited phase space available in the decay, we only consider the states with angular momenta up to 2.
As discussed in Ref.~\cite{Zou}, we define the spin projection operator
$P^{(S)}_{\mu_{1} \cdot \cdot \cdot \mu_{S}\nu_{1} \cdot \cdot \cdot \nu_{S}}$ for a process
$a \rightarrow bc$ as
\begin{eqnarray}
\begin{aligned}
P^{(1)}_{\mu \nu} = -g_{\mu\nu} + \frac{p_{a\mu}p_{a\nu}}{p^{2}_{a}}
\end{aligned}
\end{eqnarray}
for spin 1,
\begin{eqnarray}
\begin{aligned} 
&P^{(2)}_{\mu_{1}\mu_{2}\nu_{1}\nu_{2}} = \\
&\frac{1}{2}(P^{(1)}_{\mu_{1}\nu_{1}}P^{(1)}_{\mu_{2}\nu_{2}}+
P^{(1)}_{\mu_{1}\nu_{2}}P^{(1)}_{\mu_{2}\nu_{1}})
-\frac{1}{3}P^{(1)}_{\mu_{1}\mu_{2}}P^{(1)}_{\nu_{1}\nu_{2}}
\end{aligned}
\end{eqnarray}
for spin 2.
The covariant tensors $\tilde{t}^{L}_{\mu_{1}\cdot \cdot \cdot \mu_{l}}$ for the final states of pure orbital
angular momentum $L$ are constructed from relevant momenta $p_{a}$, $p_{b}$, $p_{c}$~\cite{Zou}
\begin{eqnarray}
\begin{aligned}
\tilde{t}^{L}_{\mu_{1}\cdot \cdot \cdot \mu_{L}} = 
(-1)^{L}P^{(L)}_{\mu_{1} \cdot \cdot \cdot \mu_{L}\nu_{1} \cdot \cdot \cdot \nu_{L}}
r^{\nu_{1}}\cdot \cdot \cdot r^{\nu_{L}},
\end{aligned}
\end{eqnarray}
where $r = p_{b} - p_{c}$.

Ten kinds of decay modes used in the analysis are listed in Table~\ref{tab:spin}.
We use $\tilde{T}^{(L)}_{\mu_{1}...\mu_{L}}$ to represent the
decay from the $D$ meson and $\tilde{t}^{(L)}_{\mu_{1}...\mu_{L}}$
to represent the decay from the intermediate state.

\begin{table*}[hbtp]
\begin{center}
\renewcommand{\tabcolsep}{0.86cm}
\caption{Spin factors $S(p)$ for different decay modes.}
\begin{tabular}{lc} \hline
Decay mode & $S(p)$ \\ \hline
$D[S]\rightarrow \mbox{\,V}_{1} \mbox{\,V}_{2}$, $\mbox{\,V}_{1} \rightarrow \mbox{\,P}_{1} \mbox{\,P}_{2}$,
$\mbox{\,V}_{2} \rightarrow \mbox{\,P}_{3} \mbox{\,P}_{4}$
& $\tilde{t}^{(1)\mu}(\mbox{\,V}_{1})\tilde{t}^{(1)}_{\mu}(\mbox{\,V}_{2})$ \\
$D[P]\rightarrow \mbox{\,V}_{1} \mbox{\,V}_{2}$, $\mbox{\,V}_{1} \rightarrow \mbox{\,P}_{1} \mbox{\,P}_{2}$,
$\mbox{\,V}_{2} \rightarrow \mbox{\,P}_{3} \mbox{\,P}_{4}$
& $\epsilon_{\mu\nu\lambda\sigma}p^{\mu}(D)\tilde{T}^{(1)\nu}(D)\tilde{t}^{(1)\lambda}(\mbox{\,V}_{1})\tilde{t}^{(1)\sigma}(\mbox{\,V}_{2})$\\
$D[D]\rightarrow \mbox{\,V}_{1} \mbox{\,V}_{2}$, $\mbox{\,V}_{1} \rightarrow \mbox{\,P}_{1} \mbox{\,P}_{2}$,
$\mbox{\,V}_{2} \rightarrow \mbox{\,P}_{3} \mbox{\,P}_{4}$
& $\tilde{T}^{(2)\mu\nu}(D)\tilde{t}^{(1)}_{\mu}(\mbox{\,V}_{1})\tilde{t}^{(1)}_{\nu}(\mbox{\,V}_{2})$\\
$D\rightarrow \mbox{\,AP}_{1}, \mbox{\,A}[S]\rightarrow \mbox{\,VP}_{2}$,
$\mbox{\,V} \rightarrow \mbox{\,P}_{3} \mbox{\,P}_{4}$
&$\tilde{T}_{1}^{\mu}(D)P_{\mu\nu}^{(1)}(\mbox{\,A})\tilde{t}^{(1)\nu}(\mbox{\,V})$\\
$D\rightarrow \mbox{\,AP}_{1}, \mbox{\,A}[D]\rightarrow \mbox{\,VP}_{2}$,
$\mbox{\,V} \rightarrow \mbox{\,P}_{3} \mbox{\,P}_{4}$
& $\tilde{T}^{(1)\mu}(D)\tilde{t}_{\mu\nu}^{(2)}(\mbox{\,A})\tilde{t}^{(1)\nu}(\mbox{\,V})$\\
$D \rightarrow \mbox{\,AP}_{1}, \mbox{\,A} \rightarrow \mbox{\,SP}_{2} $,
$\mbox{\,S} \rightarrow \mbox{\,P}_{3} \mbox{\,P}_{4}$
& $\tilde{T}^{(1)\mu}(D)\tilde{t}^{(1)}_{\mu}(\mbox{\,A})$\\
$D \rightarrow \mbox{\,VS}$, $\mbox{\,V} \rightarrow \mbox{\,P}_{1} \mbox{\,P}_{2}$,
$\mbox{\,S} \rightarrow \mbox{\,P}_{3} \mbox{\,P}_{4}$
& $\tilde{T}^{(1)\mu}(D)\tilde{t}^{(1)}_{\mu}(\mbox{\,V})$\\
$D \rightarrow \mbox{\,V}_{1}\mbox{\,P}_{1}, \mbox{\,V}_{1} \rightarrow \mbox{\,V}_{2}\mbox{\,P}_{2}$,
$\mbox{\,V}_{2} \rightarrow \mbox{\,P}_{3} \mbox{\,P}_{4}$
&$\epsilon_{\mu\nu\lambda\sigma}p_{\mbox{\,V}_{1}}^{\mu}q_{\mbox{\,V}_{1}}^{\nu}p_{\mbox{\,P}_{1}}^{\lambda}q_{\mbox{\,V}_{2}}^{\sigma}$\\
$D \rightarrow \mbox{\,PP}_{1}, \mbox{\,P} \rightarrow \mbox{\,VP}_{2}$,
$\mbox{\,V} \rightarrow \mbox{\,P}_{3} \mbox{\,P}_{4}$
& $p^{\mu}(\mbox{\,P}_{2})\tilde{t}^{(1)}_{\mu}(\mbox{\,V}) $\\
$D \rightarrow \mbox{\,TS}$, $\mbox{\,T} \rightarrow \mbox{\,P}_{1} \mbox{\,P}_{2}$,
$\mbox{\,S} \rightarrow \mbox{\,P}_{3} \mbox{\,P}_{4}$
& $\tilde{T}^{(2)\mu\nu}(D)\tilde{t}^{(2)}_{\mu\nu}(\mbox{\,T})$ \\
\hline
\end{tabular}
\label{tab:spin}
\end{center}
\end{table*}

\subsubsection{Blatt-Weisskopf barrier factors}
The Blatt-Weisskopf barrier factor~\cite{Blatt} $F_{L}(p_{j})$ is
a function of the angular momentum $L$ and the four-momenta
$p_{j}$ of the daughter particles.
For a process $a \rightarrow bc$,
the magnitude of the momentum $q$ of the daughter $b$ or $c$ in the rest system of $a$ is given by
\begin{eqnarray}
\begin{aligned}
\label{Qabc}
q = \sqrt{\frac{(s_{a}+s_{b}-s_{c})^2}{4s_{a}}-s_{b}}
\end{aligned}
\end{eqnarray}
with $s_{\beta} = E_{\beta}^{2} - \vec p_{\beta}^{2}, \beta=a,b,c$.
The Blatt-Weisskopf barrier factor is then given by
\begin{eqnarray}
\begin{aligned}
F_{L}(q) = z^{L}X_{L}(q), 
\end{aligned}
\end{eqnarray}
where $z = qR$. $R$ is the effective radius of the barrier, which is fixed to
$3.0{\mbox{\,GeV}^{-1}}$ for intermediate resonances and
$5.0{\mbox{\,GeV}^{-1}}$ for the $D^{0}$ meson.
$X_{L}(q)$ is given by
\begin{eqnarray}
\begin{aligned}
X_{L = 0}(q) = 1,
\end{aligned}
\end{eqnarray}
\begin{eqnarray}
\begin{aligned}
X_{L = 1}(q) = \sqrt{\frac{2}{z^{2}+1}},
\end{aligned}
\end{eqnarray}
\begin{eqnarray}
\begin{aligned}
X_{L = 2}(q) = \sqrt{\frac{13}{z^{4}+3z^{2}+9}}.
\end{aligned}
\end{eqnarray}

\subsubsection{Propagator}
The resonances $\bar{K}^{*0}$ and $a_{1}^{+}(1260)$ are parametrized
as relativistic Breit-Wigner function with
a mass depended width
\begin{eqnarray}
\begin{aligned}
P(m) = \frac{1}{(m^{2}_{0} - s_{a})-im_{0}\Gamma (m)},
\end{aligned}
\end{eqnarray}
where $m_{0}$ is the mass of resonance to be determined. $\Gamma (m)$ is given by
\begin{eqnarray}
\begin{aligned}
\label{Gamma_m}
\Gamma (m) = \Gamma_{0}\left(\frac{q}{q_{0}}\right)^{2L+1}\left(\frac{m_{0}}{m}\right)\left(\frac{X_{L}(q)}{X_{L}(q_{0})}\right)^{2},
\end{aligned}
\end{eqnarray}
where $q_{0}$ denotes the value of $q$ at $m=m_{0}$.
The $K_{1}^{-}(1270)$ is parametrized as a relativistic Breit-Wigner function with a constant width
$\Gamma (m) = \Gamma_{0}$, and the $\rho^{0}$ is parametrized with the Gounaris-Sakurai line shape~\cite{GS},
which is given by
\begin{eqnarray}
\begin{aligned}
P_\text{GS}(m) = \frac{1+d\frac{\Gamma_{0}}{m_{0}}}{(m_{0}^{2}-m^{2})+f(m)-im_{0}\Gamma (m)},
\end{aligned}
\end{eqnarray}
where
\begin{eqnarray}
\begin{aligned}
f(m) = \Gamma_{0}\frac{m_{0}^{2}}{q_{0}^{3}}\Big[q^{2}(h(m)-h(m_{0}))\\
+(m_{0}^{2}-m^{2})q_{0}^{2}\frac{dh}{d(m^{2})}\Big|_{m^{2} = m^{2}_{0}}\Big],&
\end{aligned}
\end{eqnarray}
and the function $h(m)$ is defined as
\begin{eqnarray}
\begin{aligned}
h(m) = \frac{2}{\pi}\frac{q}{m}\ln\left(\frac{m+2q}{2m_{\pi}}\right),
\end{aligned}
\end{eqnarray}
with
\begin{eqnarray}
\begin{aligned}
\frac{dh}{d(m^{2})}\Big|_{m^{2} = m^{2}_{0}} = h(m_{0})[(8q_{0}^{2})^{-1}-(2m_{0}^{2})^{-1}]+(2\pi m_{0}^{2})^{-1},
\end{aligned}
\end{eqnarray}
where $m_{\pi}$ is the charged pion mass.
The normalization condition at $P_\text{GS}(0)$ fixes the parameter $d=f(0)/(\Gamma_{0}m_{0})$.
It is found to be~\cite{GS}
\begin{eqnarray}
\begin{aligned}
d = \frac{3}{\pi}\frac{m_{\pi}^{2}}{q_{0}^{2}}\ln\left(\frac{m_{0}+2q_{0}}{2m_{\pi}}\right)+\frac{m_{0}}{2\pi q_{0}}
- \frac{m_{\pi}^{2}m_{0}}{\pi q_{0}^{3}}.
\end{aligned}
\end{eqnarray}

\subsubsection{Parametrization of the $K\pi$ $S$-wave}
For the $K\pi$ $S$-wave [denoted as $(K\pi)_{{\rm S-wave}}$], we use the same parametrization as $BABAR$~\cite{KPiS},
which is extracted from scattering data~\cite{LASS}.
The model is built from a Breit-Wigner shape for the $\bar{K}^{*}_{0}(1430)^0$ combined with an effective
range parametrization for the nonresonant component given by
\begin{eqnarray}
\begin{aligned}
A(m_{K\pi}) = F\sin\delta_{F}e^{i\delta_{F}} + R\sin\delta_{R}e^{i\delta_{R}}e^{i2\delta_{F}},
\end{aligned}
\end{eqnarray}
with
\begin{align}
\delta_{F} &= \phi_{F} +
             \cot^{-1}\left[\frac{1}{aq}+\frac{rq}{2}\right], \\
\delta_{R} &= \phi_{R} + \tan^{-1}\left[\frac{M\Gamma(m_{K\pi})}{M^{2}-m_{K\pi}^{2}}\right],
\end{align}
where $a$ and $r$ denote the scattering length and effective interaction length.
$F~(\phi_{F})$ and $R~(\phi_{R})$ are the relative magnitudes (phases) for
the nonresonant and resonant terms, respectively.
$q$ and $\Gamma(m_{K\pi})$ are defined as in Eq.~(\ref{Qabc}) and Eq.~(\ref{Gamma_m}), respectively.
In the fit, the parameters $M$, $\Gamma$, $F$, $\phi_{F}$, $R$, $\phi_{R}$, $a$ and $r$
are fixed to the values obtained from the fit to the
$D^{0} \rightarrow K_{S}^{0}\pi^{+}\pi^{-}$ Dalitz plot~\cite{KPiS},
as summarized in Table ~\ref{tab:BABAR KPiS}. 
These fixed parameters will be varied within their uncertainties to estimate the corresponding 
systematic uncertainties, which is discussed in detail in Sec.~\ref{Model_sys}.
\begin{table}[htp]
\begin{center}
\caption{$K\pi$ $S$-wave parameters,
obtained from the fit to the $D^{0} \rightarrow K_{S}^{0}\pi^{+}\pi^{-}$
Dalitz plot from $BABAR$~\cite{KPiS}.}
\begin{tabular}{cc} \hline
$M$(GeV/$c^{2}$) & $1.463\pm0.002$ \\
$\Gamma$(GeV/$c^{2}$) & $0.233\pm0.005$ \\
$F$ & $0.80\pm0.09$ \\
$\phi_{F}$ & $2.33\pm0.13$ \\
$R$ & $1$(fixed)\\
$\phi_{R}$ & $-5.31\pm0.04$ \\
$a$ & $1.07\pm0.11$ \\
$r$ & $-1.8\pm0.3$ \\
\hline
\end{tabular}
\label{tab:BABAR KPiS}
\end{center}
\end{table}

\subsection{Fit fraction and the statistical uncertainty}
\label{FF_cal}
We divide the fit model into several components according to the intermediate resonances,
which can be found in Sec.~\ref{Nominal Fit}.
The fit fractions of the individual components (amplitudes) are calculated according to the
fit results and are compared to other measurements.
In the calculation, a large phase space (PHSP) MC sample with neither detector acceptance nor resolution involved is used.
The fit fraction for an amplitude or a component (a certain subset of amplitudes) is defined as
\begin{eqnarray}
\begin{aligned}
\label{eq_FF}
FF(n)  = \frac{\sum_{k=1}^{N_{{\rm gen}}}|\tilde{A}_{\bf n}(p_{j}^{k})|^{2}}
{\sum_{k=1}^{N_{{\rm gen}}}|M(p_{j}^{k})|^{2}},
\end{aligned}
\end{eqnarray}
where $\tilde{A}_{\bf n}(p_{j}^{k})$ is either the $n^{{\rm th}}$ amplitude
[$\tilde{A}_{{\bf n}}(p_{j}^{k}) = c_{n}A_{n}(p_{j}^{k})$]
or the ${\bf n}^{{\rm th}}$ component of a coherent sum of amplitudes
[$\tilde{A}_{{\bf n}}(p_{j}^{k}) = \sum{c_{n_{i}}A_{n_{i}}(p_{j}^{k})}$],
$N_{{\rm gen}}$ is the number of the PHSP MC events.

To estimate the statistical uncertainties of the fit fractions,
we repeat the calculation of fit fractions by
randomly varying the fitted parameters according to the error matrix.
Then, for every amplitude or component, we fit the resulting distribution with a Gaussian function,
whose width gives the corresponding statistical uncertainty.

\subsection{Goodness of fit}
\label{Goodness of Fit}
To examine the performance of the fit process, the goodness of fit is
defined as follows.
Since the $D^{0}$ and all four final states particles have spin zero, the phase space of the decay
$D^{0} \rightarrow K^{-}\pi^{+}\pi^{+}\pi^{-}$ can be completely described by five
linearly independent Lorentz invariant variables.
Denoting as $\pi^+_{1}$ the one of the two identical pions which results in a higher
$\pi^{+}\pi^{-}$ invariant mass
and the other pion as $\pi^+_{2}$, we choose the five invariant masses
$m_{\pi^{+}_{1}\pi^{-}}$, $m_{\pi^{+}_{2}\pi^{-}}$,
$m_{K^{-}\pi^{+}_{1}\pi^{-}}$, $m_{\pi^{+}_{1}\pi^{+}_{2}\pi^{-}}$ and
$m_{K^{-}\pi^{+}_{1}\pi^{+}_{2}}$.
To calculate the goodness of fit, the five-dimensional phase space is
first divided into cells with
equal size. Then, adjacent cells are combined until the number of events in each cell is larger than 20.
The deviation of the fit in each cell is calculated, $\chi_{p} = \frac{N_{p} - N_{p}^{{\rm exp}}}{\sqrt{N_{p}^{{\rm exp}}}}$,
and the goodness of fit is quantified as $\chi^{2} = \sum_{p=1}^{n}\chi_{p}^{2}$,
where $N_{p}$ and $N_{p}^{{\rm exp}}$ are the number of the observed events and
the expected number determined from the fit results in the $p^{th}$ cell, respectively,
and $n$ is the total number of cells.
The number of degrees of freedom (NDF) $\nu$ is given by $\nu = (n-1) - n_{{\rm par}}$, where $n_{{\rm par}}$
is the number of the free parameters in the fit.

\section{Results}
\label{RESULTS}
\label{Nominal Fit}
In order to determine the optimal set of amplitude that contribute to 
the decay $D^{0} \rightarrow K^{-} \pi^{+} \pi^{+} \pi^{-}$,  
considering the results in PDG~\cite{PDG}, 
we start with the fit including the components with significant contribution 
and add more amplitude in the fit one by one. 
The corresponding statistical significance for the new amplitude is calculated with the change of the 
log-likelihood value $\Delta{\ln{L}}$, taking the change of the degrees of
freedom $\Delta \nu$ into account.

In the $K^{-}\pi^{+}$ and $\pi^{+}\pi^{-}$ invariant mass spectra, there are 
clear structures for $\bar{K}^{*0}$ and $\rho^{0}$. 
The intermediate resonance $K_{1}^{-}(1270)$ is observed
with $K_{1}^{-}(1270)\rightarrow \bar{K}^{*0}\pi^{-}$ or $K^{-}\rho^{0}$.
In the $\pi^{+}\pi^{+}\pi^{-}$ invariant mass spectrum, a broad bump appears.  
We find this bump can be fitted as $a_{1}^{+}(1260)$, which was also
observed by the Mark III~\cite{MarKIII} experiment.
If it is fitted with a nonresonant $(\rho^{0} \pi^{+})_{{\rm A}}$ amplitude instead, we find that the significance for 
$a_{1}^{+}(1260)$ with respect to $(\rho^{0} \pi^{+})_{{\rm A}}$ is larger than $10\sigma$. 
The three-body nonresonant states come from two kinds of 
contributions, $K^{-}\pi^{+}\rho^{0}$ and $\bar{K}^{*0}\pi^{+}\pi^{-}$. 
The $\bar{K}^{*0}\pi^{-} \slash K^{-}\rho^{0}$ 
can be in a pseudoscalar, a vector or an axial-vector state, while  
the $K^{-}\pi^{+} \slash \pi^{+}\pi^{-}$ can be in a scalar state. 
The four-body nonresonant states are relatively complex, such as  
$D \rightarrow$ VV, $D \rightarrow$ VS, $D \rightarrow$ TS, $D \rightarrow$ TV, 
$D \rightarrow$ AP with $\mbox{\,A} \rightarrow$ VP or SP, all of which may contribute to the decay. 
Since the process $D^{0} \rightarrow K^{-}a_{1}^{+}(1260)$,
$a_{1}^{+}(1260)[S] \rightarrow \rho^{0} \pi^{+}$ has the largest fit fraction,
we fix the corresponding magnitude and phase to $1.0$ and $0.0$ and allow the
magnitudes and phases of the other processes to vary in the fit.

We keep the processes with significance larger than $5\sigma$ for the next iteration. 
The fit involving both the $K^{-}a_{1}^{+}(1260)$
and the nonresonant $K^{-}(\rho^{0}\pi^{+})_{{\rm A}}$ contribution
does not result in a significantly improvement of fit, 
but the fit fractions of the two amplitudes are much different
with the assumption of only $K^{-}a_{1}^{+}(1260)$ and are nearly 100\% correlated.
We avoid this kind of case and only consider the resonant term, 
in agreement with the analysis of Mark III~\cite{MarKIII}.
For the process $D^{0} \rightarrow K^{-}_{1}(1270)\pi^{+}$ with 
$K^{-}_{1}(1270)[S] \rightarrow \bar{K}^{*0}\pi^{-}$, 
the corresponding significance is found to be $4.3\sigma$ only,
but we still include it in the fit 
since the corresponding $D$-wave process is found to have 
a statistical significance of larger than $9\sigma$.
Better projections in the invariant mass spectra
and an improved fit quality $\chi^{2}$ are also seen with this $S$-wave process included.

\begin{figure*}[hbtp]
\begin{center}
\begin{minipage}[b]{0.35\textwidth}
\epsfig{width=1.00\textwidth,clip=true,file=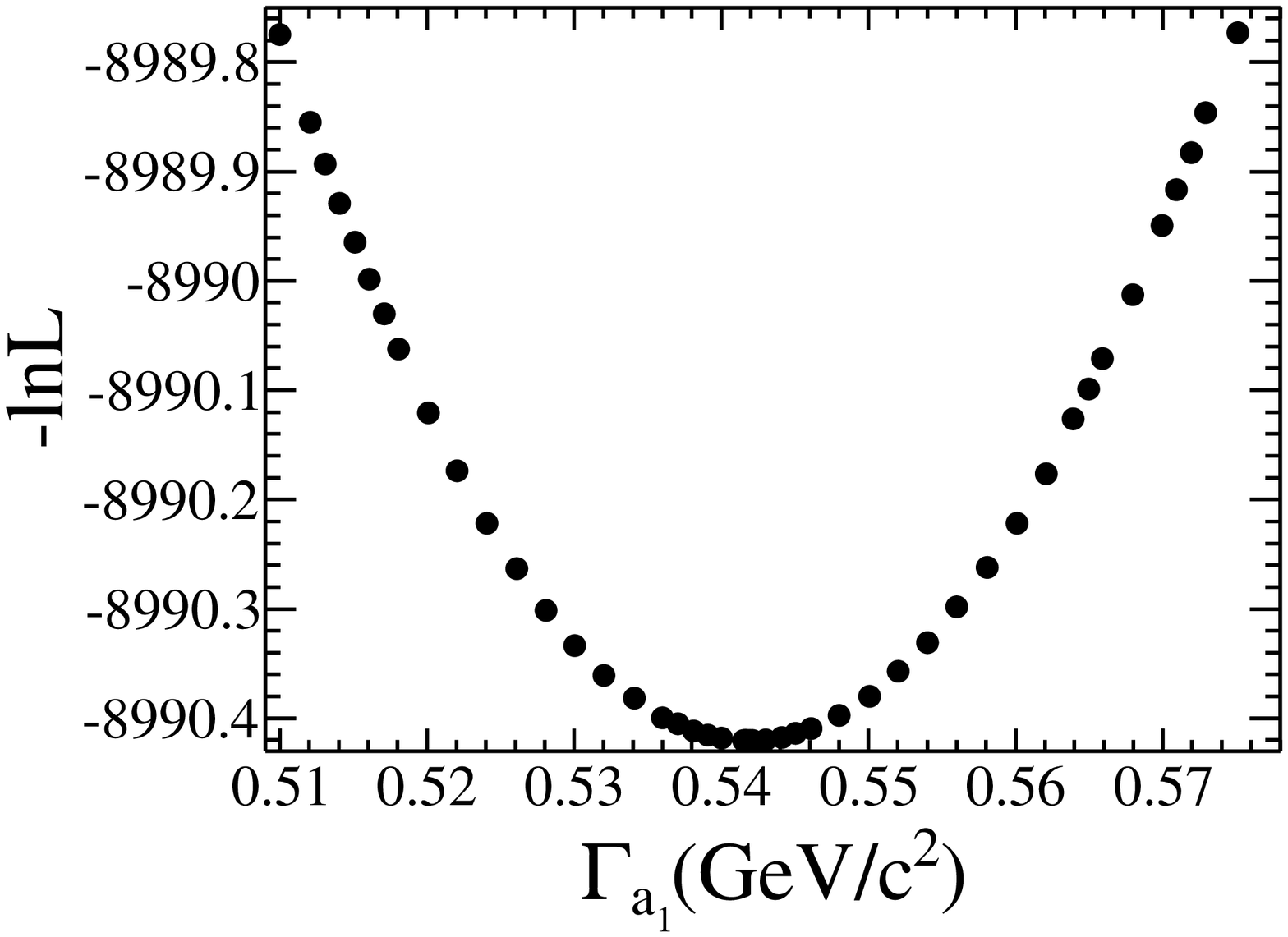}
\put(-70,85){(a)}
\end{minipage}
\begin{minipage}[b]{0.35\textwidth}
\epsfig{width=1.00\textwidth,clip=true,file=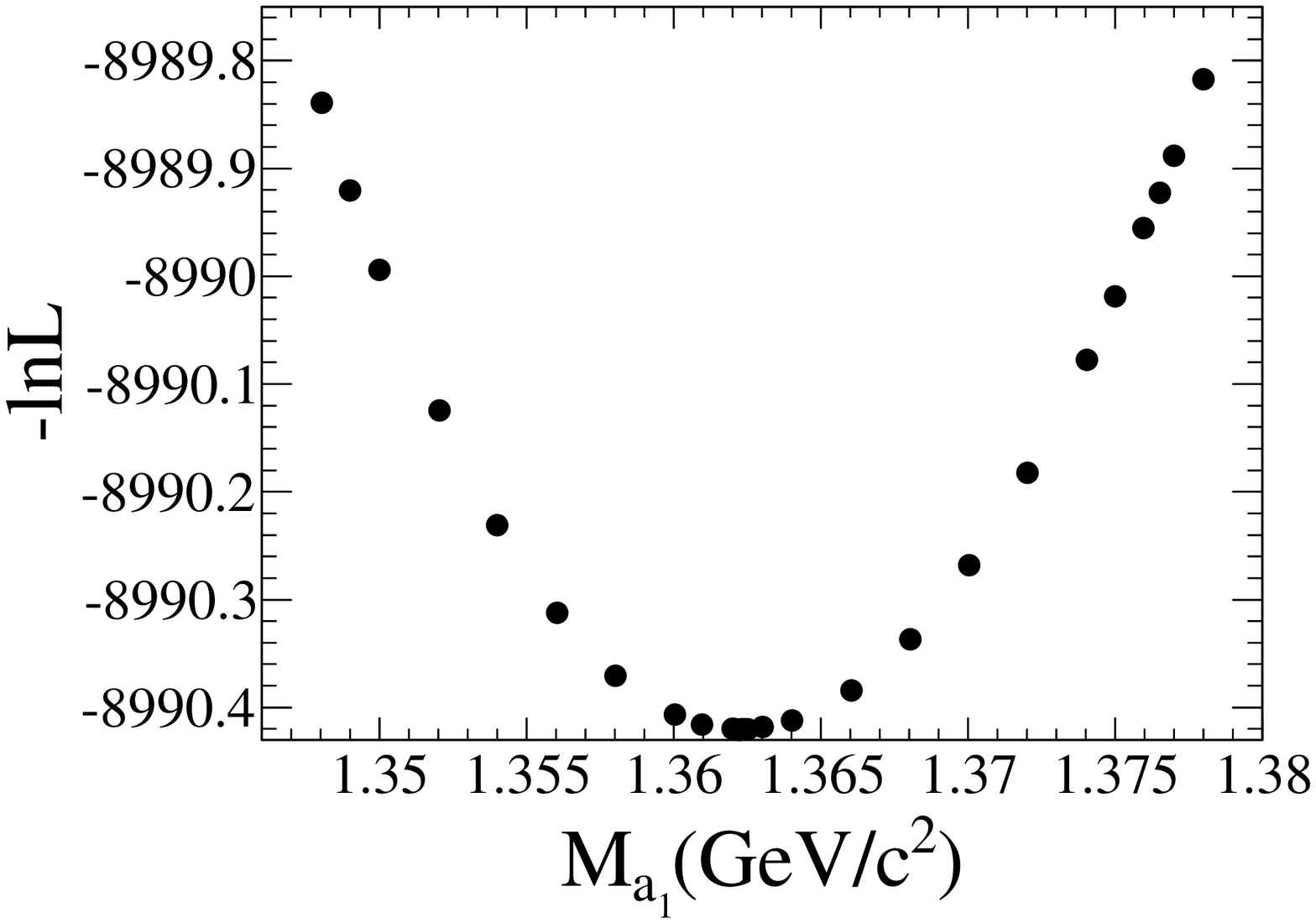}
\put(-70,85){(b)}
\end{minipage}
\caption{Likelihood scans of the width (a) and mass (b) of $a_{1}^{+}(1260)$.}
\label{fig:scan}
\end{center}
\end{figure*}

Finally, we retain 23 processes categorized into seven components.
The other processes, not used in our nominal results but have been tested when determining 
the nominal fit model, are listed in Appendix A.
The widths and masses of $\bar{K}^{*0}$ and $\rho^{0}$ are determined by the fit.
The results of are listed in Table~\ref{tab:mass_width}.
\begin{table}[htp]
\renewcommand\arraystretch{1.5}
\begin{center}
\caption{Masses and widths of intermediate resonances $\bar{K}^{*0}$ and $\rho^{0}$,
the first and second uncertainties are statistical and systematic, respectively.}
\begin{tabular}{ccc}\hline
Resonances & Mass (MeV/$c^{2}$) & Width (MeV/$c^{2}$)\\ \hline 
$\bar{K}^{*0}$ & $894.78 \pm 0.75 \pm 1.66$ & $44.18 \pm 1.57 \pm 1.39$\\ 
$\rho^{0}$ & $779.14 \pm 1.68 \pm 3.98$ & $148.42 \pm 2.87 \pm 3.36$\\ \hline
\end{tabular}
\label{tab:mass_width}
\end{center}
\end{table}
The $K_{1}^{-}(1270)$ has a small fit fraction, and
we fix its mass and width to the PDG values~\cite{PDG}.
The $a_{1}^{+}(1260)$ has a mass close to the upper boundary of the 
$\pi^{+}\pi^{+}\pi^{-}$ invariant mass spectrum.
Therefore, we determine its mass and width with a likelihood scan,
as shown in Fig.~\ref{fig:scan}. The scan results are
\begin{eqnarray}
\begin{aligned}
m_{a_{1}^{+}(1260)} &= 1362 \pm 13 ~\mbox{\,MeV}/c^{2}, \\
\Gamma_{a_{1}^{+}(1260)} &= 542 \pm 29 ~\mbox{\,MeV}/c^{2},
\end{aligned}
\end{eqnarray}
where the uncertainties are statistical only.
The mass and width of $a_{1}^{+}(1260)$ are fixed to the scanned values in the nominal fit.

\begin{figure*}[hbtp]
\centering
\begin{minipage}[b]{0.28\textwidth}
\epsfig{width=1.00\textwidth,clip=true,file=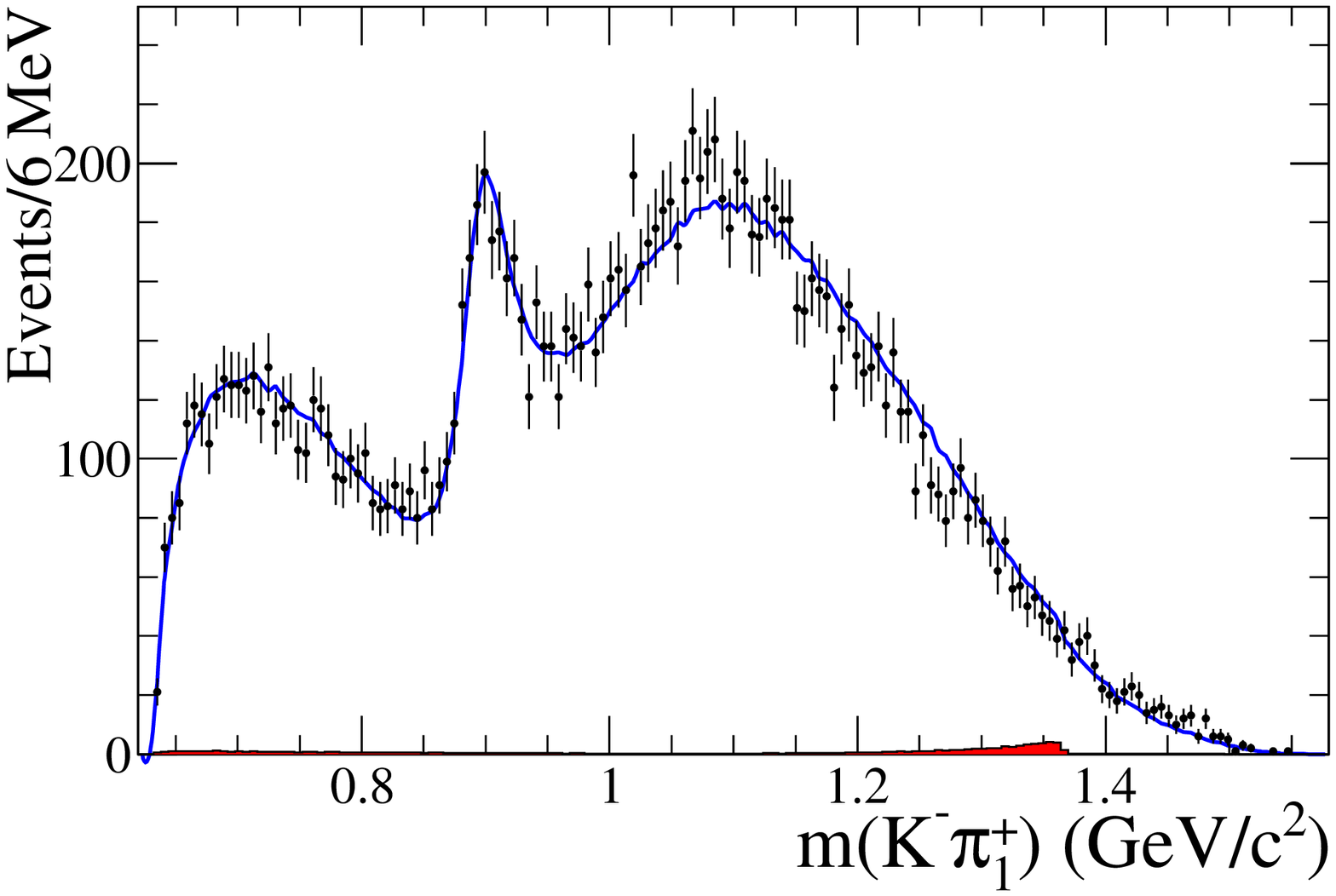}
\put(-25,75){(a)}
\end{minipage}
\begin{minipage}[b]{0.28\textwidth}
\epsfig{width=1.00\textwidth,clip=true,file=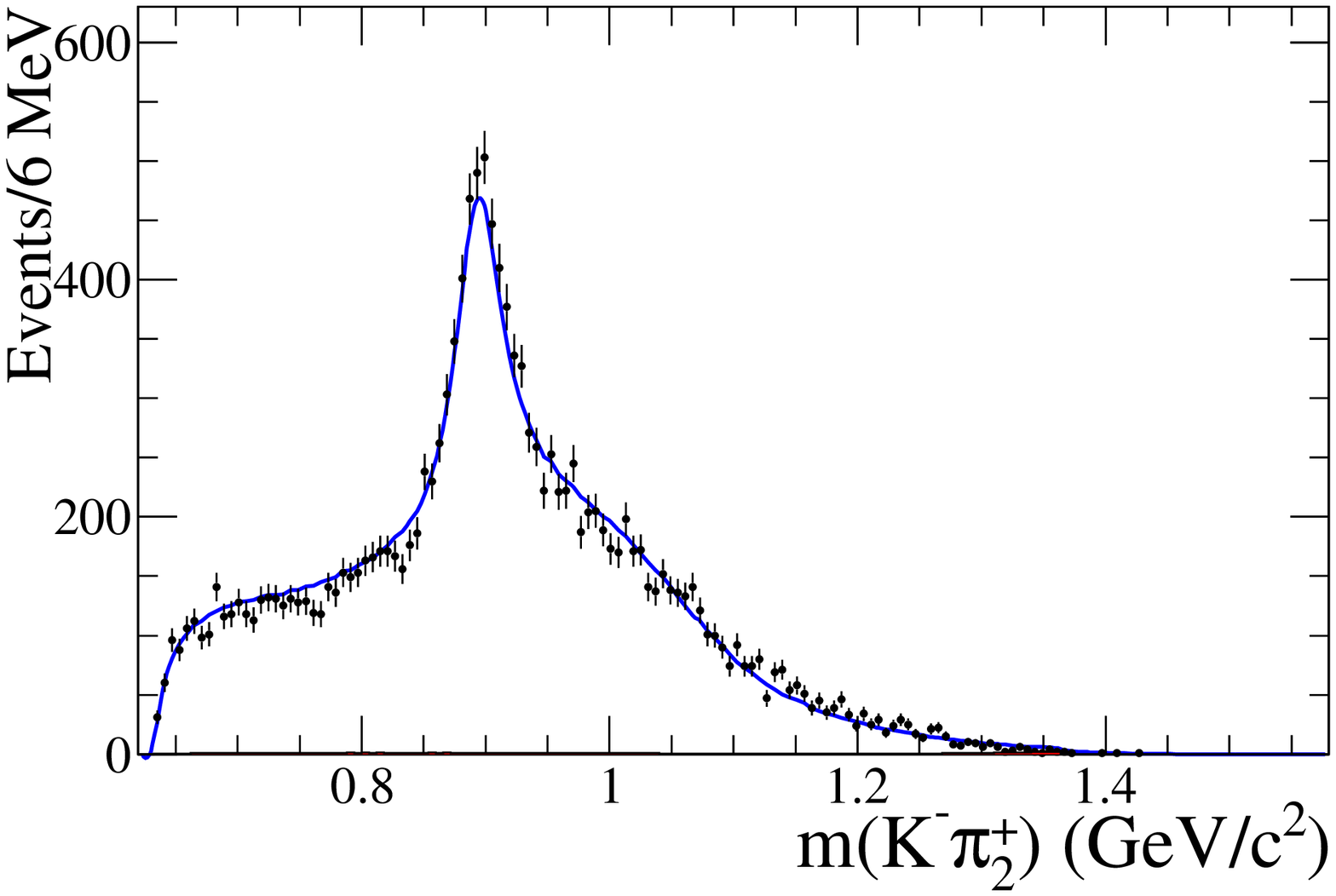}
\put(-25,75){(b)}
\end{minipage}
\begin{minipage}[b]{0.28\textwidth}
\epsfig{width=1.00\textwidth,clip=true,file=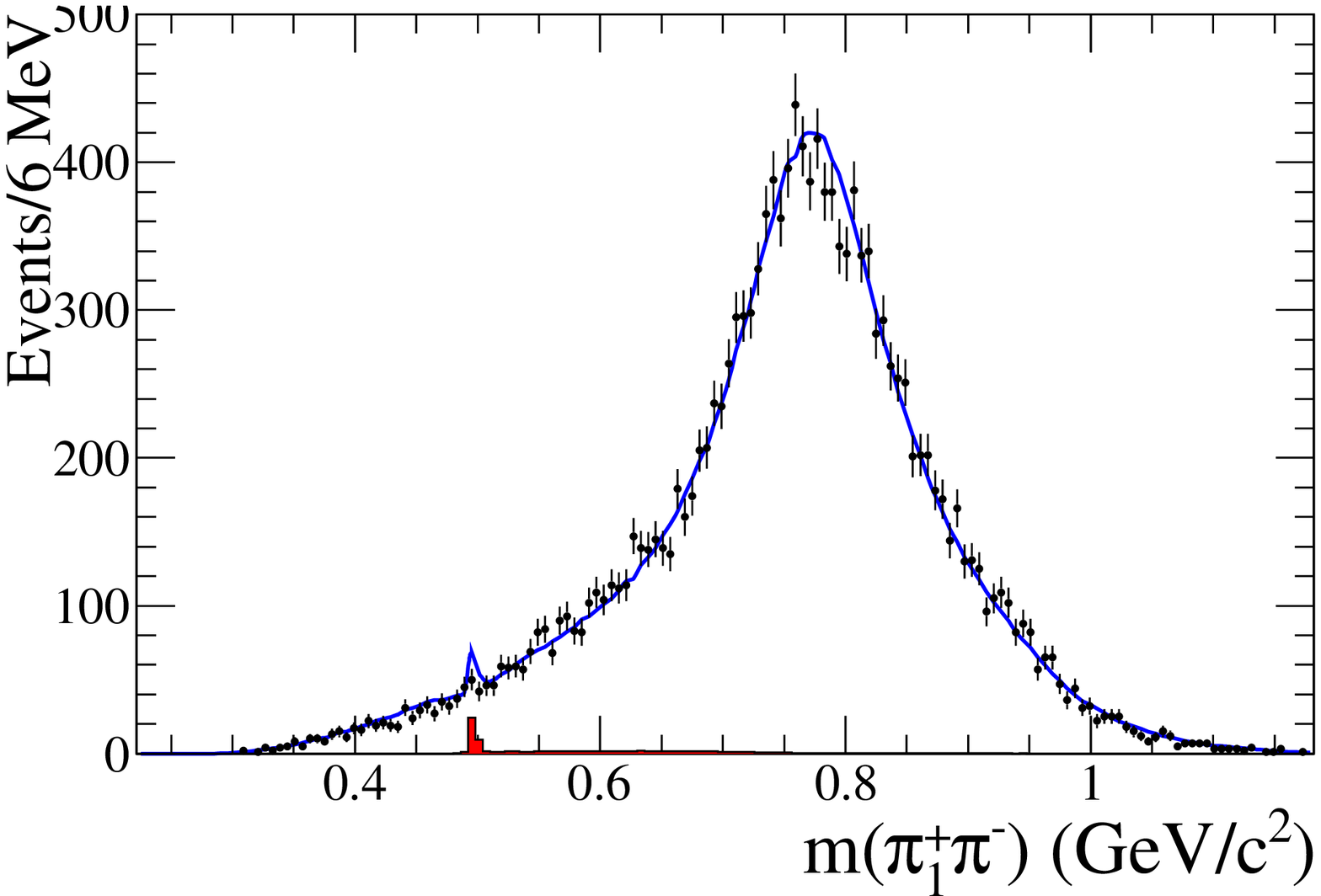}
\put(-25,75){(c)}
\end{minipage}
\begin{minipage}[b]{0.28\textwidth}
\epsfig{width=1.00\textwidth,clip=true,file=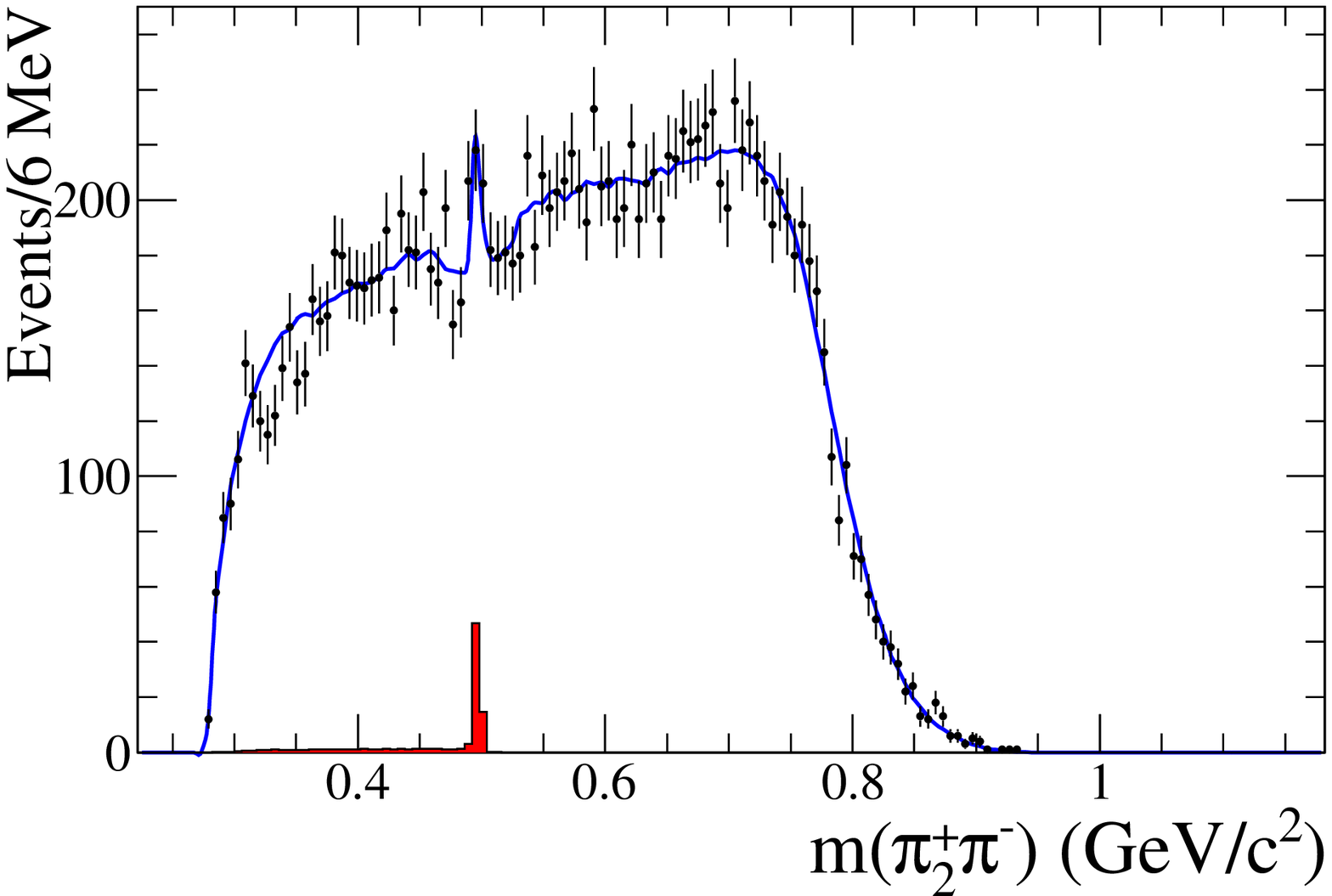}
\put(-25,75){(d)}
\end{minipage}
\begin{minipage}[b]{0.28\textwidth}
\epsfig{width=1.00\textwidth,clip=true,file=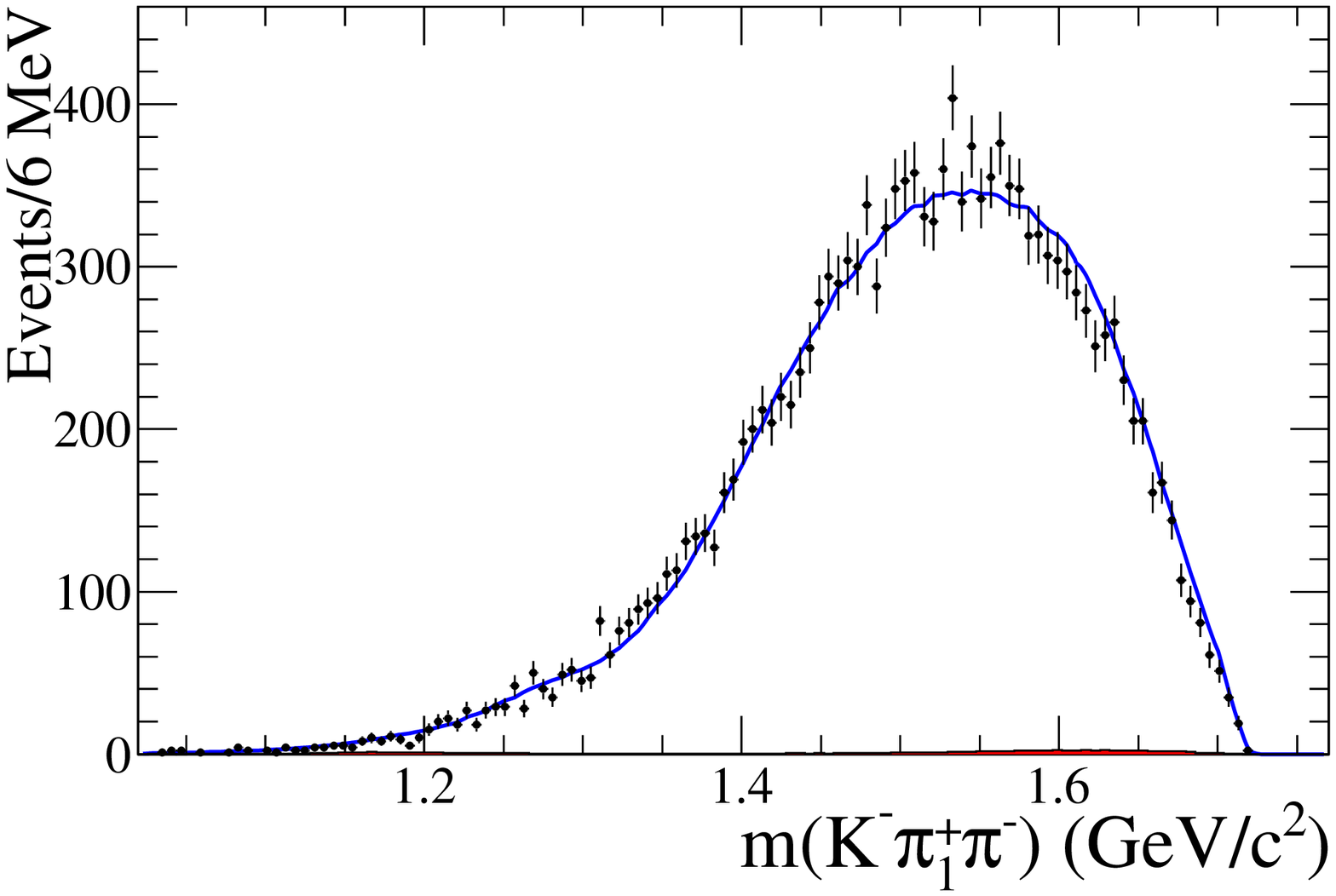}
\put(-25,75){(e)}
\end{minipage}
\begin{minipage}[b]{0.28\textwidth}
\epsfig{width=1.00\textwidth,clip=true,file=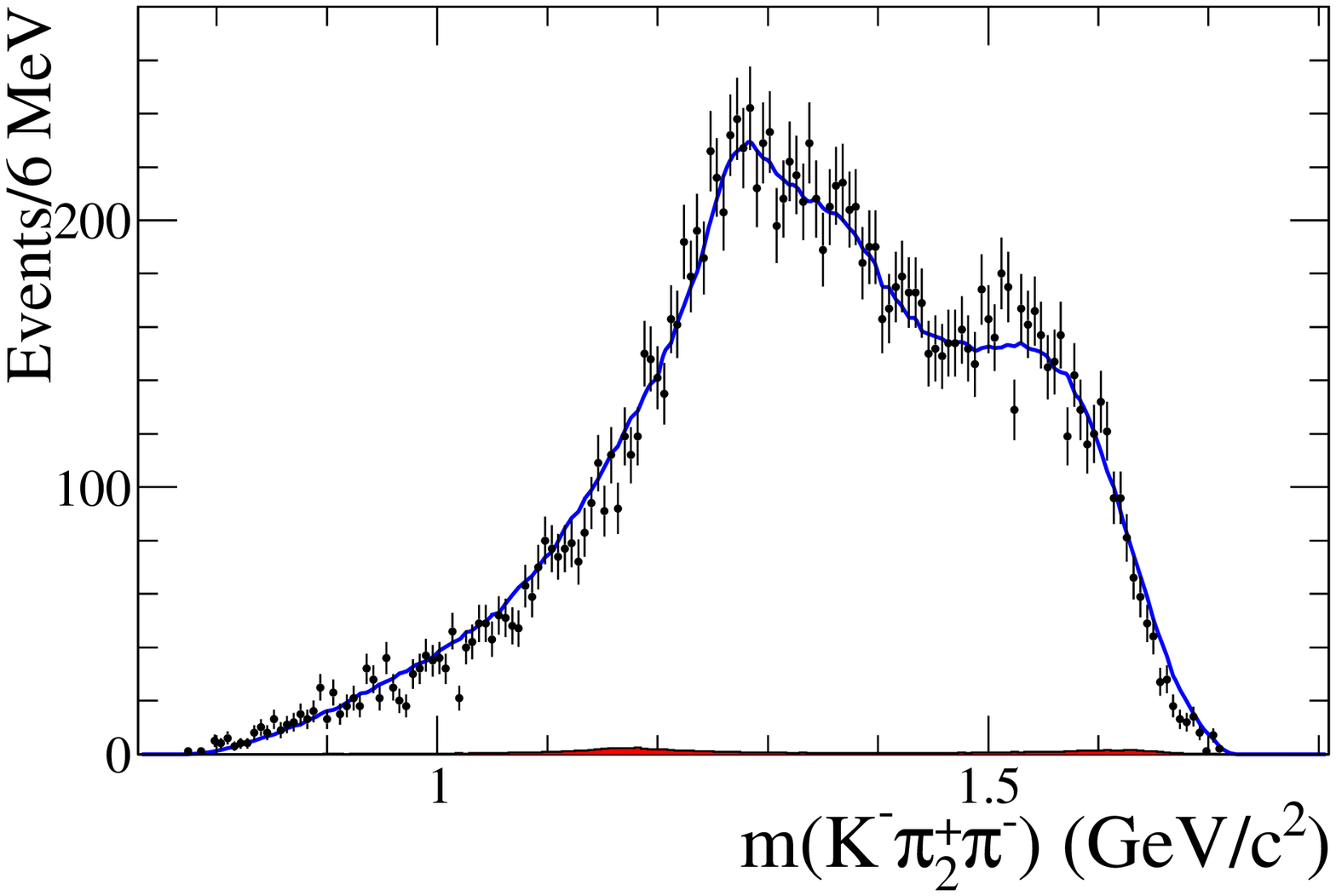}
\put(-25,75){(f)}
\end{minipage}
\begin{minipage}[b]{0.28\textwidth}
\epsfig{width=1.00\textwidth,clip=true,file=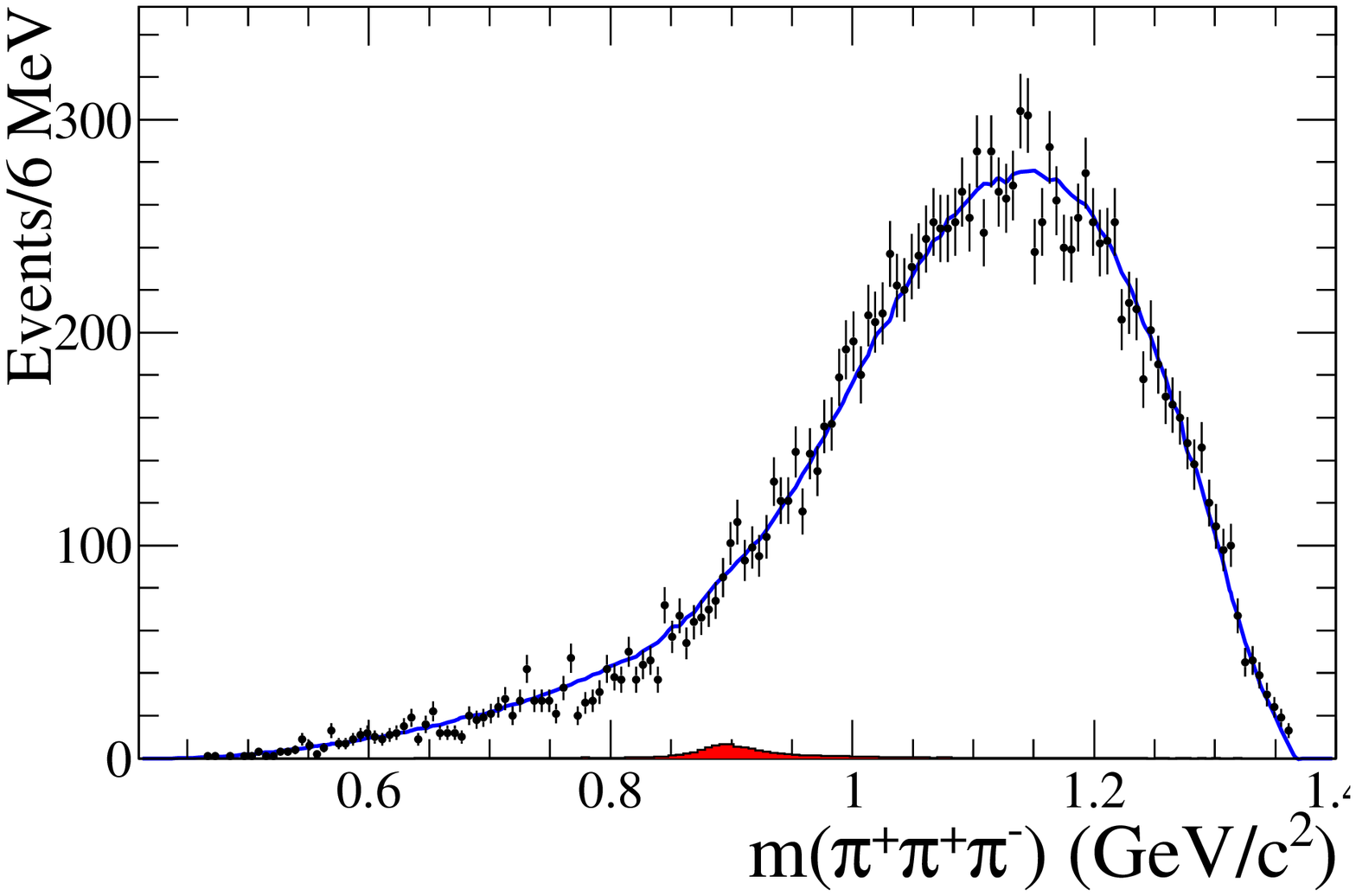}
\put(-25,75){(g)}
\end{minipage}
\begin{minipage}[b]{0.28\textwidth}
\epsfig{width=1.00\textwidth,clip=true,file=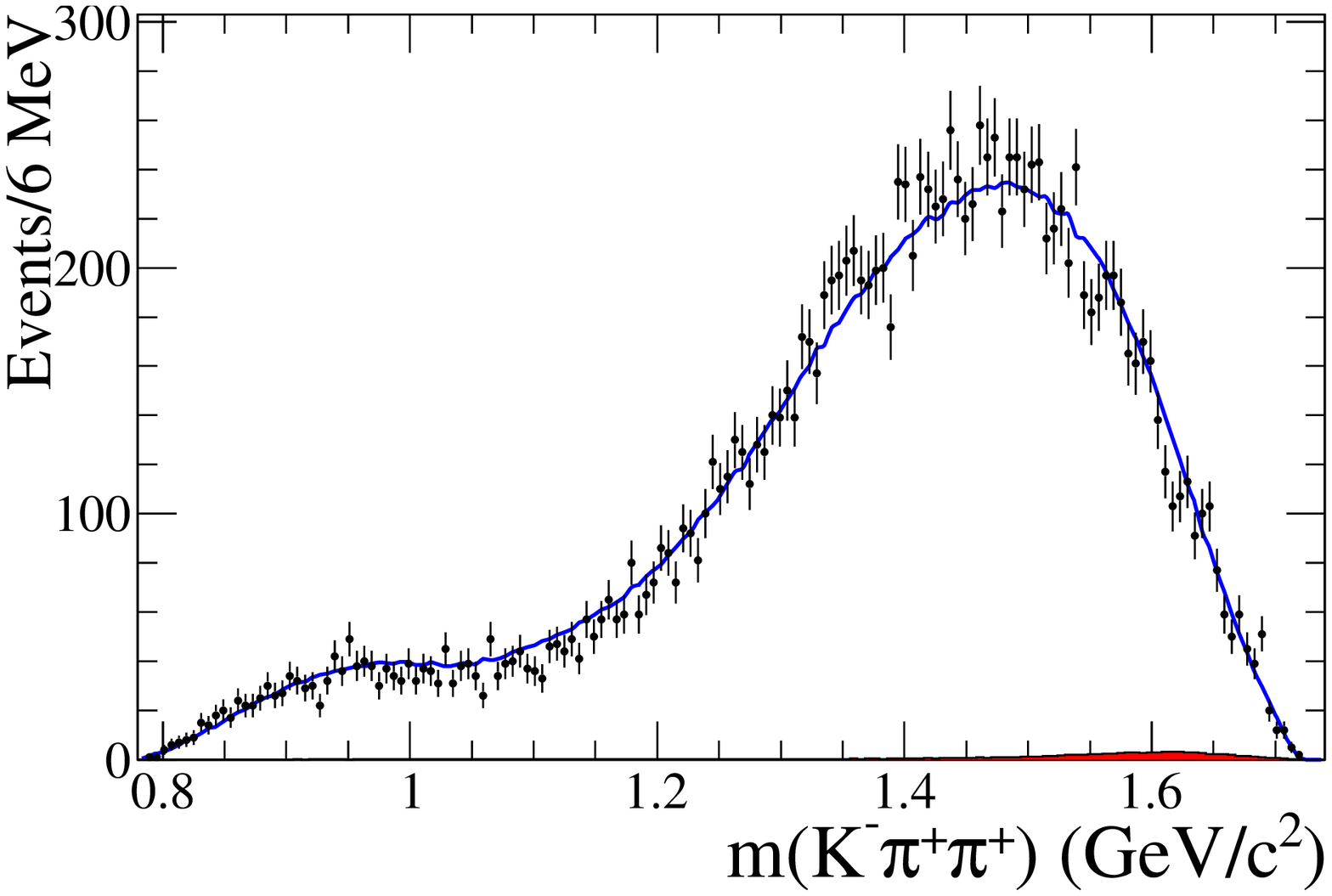}
\put(-25,75){(h)}
\end{minipage}
\begin{minipage}[b]{0.28\textwidth}
\epsfig{width=1.00\textwidth,clip=true,file=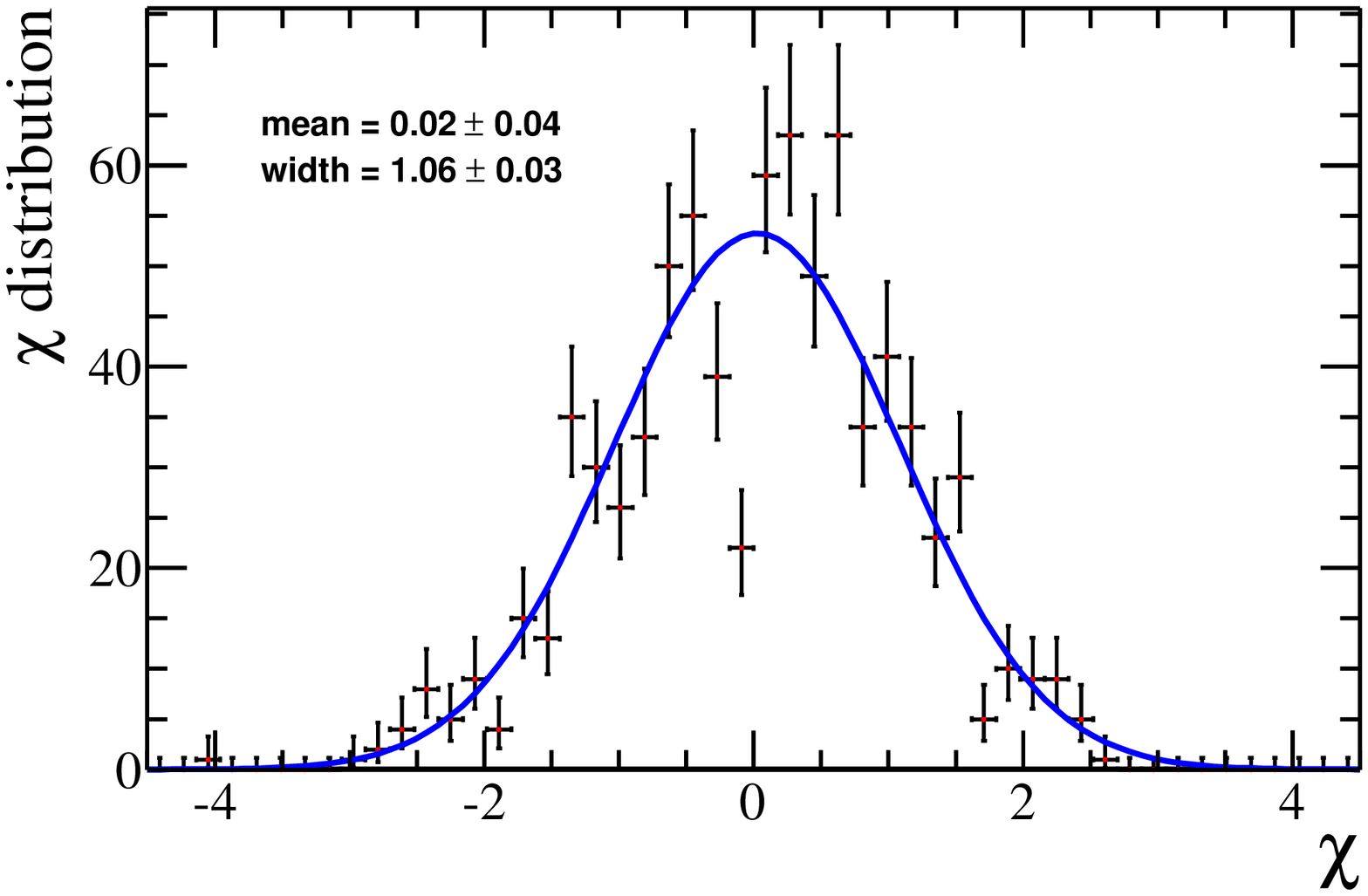}
\put(-25,75){(i)}
\end{minipage}
\caption{
Distribution of (a) $m_{K^{-}\pi^{+}_{1}}$, (b) $m_{K^{-}\pi^{+}_{2}}$, (c) $m_{\pi^{+}_{1}\pi^{-}}$, 
           (d) $m_{\pi^{+}_{2}\pi^{-}}$, (e) $m_{K^{-}\pi^{+}_{1}\pi^{-}}$, (f) $m_{K^{-}\pi^{+}_{2}\pi^{-}}$, 
           (g) $m_{\pi^{+}_{1} \pi^{+}_{2} \pi^{-}}$ and (h) $m_{K^{-}\pi^{+}_{1}\pi^{+}_{2}}$, 
where the dots with error are data, and curves are for the fit projections.
The small red histograms in each projection shows the $D^{0} \rightarrow K_{S}^{0}K^{-}\pi^{+}$ 
peaking background. In (d), a peak of $K_{S}^{0}$ can be seen, which is consistent with the MC expectation.  
The dip around the $K_{S}^{0}$ peak is caused by the requirements used to suppress the 
$D^{0} \rightarrow K_{S}^{0}K^{-}\pi^{+}$ background. 
Plot (i) shows the fit (curve) to the distribution of the
$\chi$ (points with error bars) with a Gaussian function and the fitted values of the parameters (mean and width of Gaussian).}
\label{fig:proj}
\end{figure*}

Our nominal fit yields a goodness of fit value of $\chi^{2}/\nu = 843.445 / 748 = 1.128$.
To calculate the statistical significance of a process, we repeat the fit process without the 
corresponding process included, and the changes of log-likelihood value 
and the number of free degree are taken into consideration.
The projections for eight invariant mass and the distribution of $\chi$ are shown in Fig.~\ref{fig:proj}.
All of the components, amplitudes and the significance of amplitudes are listed in Table~\ref{tab:amps}.
The fit fractions of all components are given in Table~\ref{tab:FF for comp}. 
The phases and fit fractions of all amplitudes are given in Table~\ref{tab:Phase and FF}.

\begin{table*}[hbtp]
\begin{center}
\caption{Statistical significances for different amplitudes.}
\begin{tabular}{lcc} \hline
Component&Amplitude & Significance ($\sigma$) \\ \hline
\multirow{3}{*}{
$D^{0} \rightarrow \bar{K}^{*0} \rho^{0}$}
&$D^{0}[S] \rightarrow \bar{K}^{*0} \rho^{0}$  & $>10.0$ \\
&$D^{0}[P] \rightarrow \bar{K}^{*0} \rho^{0}$  & $>10.0$ \\
&$D^{0}[D] \rightarrow \bar{K}^{*0} \rho^{0}$  & $>10.0$ \\ \hline
\multirow{2}{*}{$D^{0} \rightarrow K^{-}a_{1}^{+}(1260)$, $a_{1}^{+}(1260) \rightarrow \rho^{0}\pi^{+}$}
&$D^{0} \rightarrow K^{-}a_{1}^{+}(1260)$, $a_{1}^{+}(1260)[S] \rightarrow \rho^{0}\pi^{+}$&$>10.0$ \\
&$D^{0} \rightarrow K^{-}a_{1}^{+}(1260)$, $a_{1}^{+}(1260)[D] \rightarrow \rho^{0}\pi^{+}$&$7.4$ \\ \hline
\multirow{2}{*}{$D^{0} \rightarrow K_{1}^{-}(1270)\pi^{+}$, $K_{1}^{-}(1270) \rightarrow \bar{K}^{*0}\pi^{-}$}
&$D^{0} \rightarrow K_{1}^{-}(1270)\pi^{+}$, $K_{1}^{-}(1270)[S] \rightarrow \bar{K}^{*0}\pi^{-}$&$4.3$\\
&$D^{0} \rightarrow K_{1}^{-}(1270)\pi^{+}$, $K_{1}^{-}(1270)[D] \rightarrow \bar{K}^{*0}\pi^{-}$&$9.6$\\ \hline
$D^{0} \rightarrow K_{1}^{-}(1270)\pi^{+}$, $K_{1}^{-}(1270) \rightarrow K^{-}\rho^{0}$
&$D^{0} \rightarrow K_{1}^{-}(1270)\pi^{+}$, $K_{1}^{-}(1270)[S] \rightarrow K^{-}\rho^{0}$ & $>10.0$ \\ \hline
\multirow{4}{*}{$D^{0} \rightarrow K^{-}\pi^{+}\rho^{0}$}
&$D^{0} \rightarrow (\rho^{0} K^{-})_{{\rm A}} \pi^{+}$, 
$(\rho^{0} K^{-})_{{\rm A}}[D] \rightarrow K^{-}\rho^{0}$ & $9.6$ \\
&$D^{0} \rightarrow (K^{-}\rho^{0})_{{\rm P}}\pi^{+}$ & $7.0$ \\
&$D^{0} \rightarrow (K^{-}\pi^{+})_{{\rm S-wave}}\rho^{0}$ & $5.1$ \\
&$D^{0} \rightarrow (K^{-}\rho^{0})_{{\rm V}}\pi^{+}$ & $6.8$ \\ \hline
\multirow{3}{*}{$D^{0} \rightarrow \bar{K}^{*0}\pi^{+}\pi^{-}$}
&$D^{0} \rightarrow (\bar{K}^{*0}\pi^{-})_{{\rm P}}\pi^{+}$ & $8.5$ \\
&$D^{0} \rightarrow \bar{K}^{*0}(\pi^{+}\pi^{-})_{{\rm S}}$ & $8.9$ \\
&$D^{0} \rightarrow (\bar{K}^{*0}\pi^{-})_{{\rm V}}\pi^{+}$ & $9.7$ \\ \hline
\multirow{8}{*}{$D \rightarrow K^{-}\pi^{+}\pi^{+}\pi^{-}$}
&$D^{0} \rightarrow ((K^{-}\pi^{+})_{{\rm S-wave}}\pi^{-})_{{\rm A}}\pi^{+}$&$>10.0$ \\
&$D^{0} \rightarrow K^{-}((\pi^{+}\pi^{-})_{{\rm S}}\pi^{+})_{{\rm A}}$&$>10.0$ \\
&$D^{0} \rightarrow (K^{-}\pi^{+})_{{\rm S-wave}}(\pi^{+}\pi^{-})_{{\rm S}}$&$>10.0$ \\
&$D^{0}[S] \rightarrow (K^{-}\pi^{+})_{{\rm V}}(\pi^{+}\pi^{-})_{{\rm V}} $ & $8.8$ \\
&$D^{0} \rightarrow (K^{-}\pi^{+})_{{\rm S-wave}}(\pi^{+}\pi^{-})_{{\rm V}}$ & $5.8$ \\
&$D^{0} \rightarrow (K^{-}\pi^{+})_{{\rm V}}(\pi^{+}\pi^{-})_{{\rm S}}$ & $>10.0$ \\
&$D^{0} \rightarrow (K^{-}\pi^{+})_{{\rm T}}(\pi^{+}\pi^{-})_{{\rm S}}$ & $6.8$ \\
&$D^{0} \rightarrow (K^{-}\pi^{+})_{{\rm S-wave}}(\pi^{+}\pi^{-})_{{\rm T}}$ & $9.7$ \\ \hline
\end{tabular}
\label{tab:amps}
\end{center}
\end{table*}

\begin{table*}[hbtp]
\begin{center}
\caption{Fit fractions for different components.
The first and second uncertainties are statistical and systematic, respectively.}
\begin{tabular}{lccc} \hline
Component & Fit fraction (\%) & Mark III's result & E691's result\\ \hline
$D^{0} \rightarrow \bar{K}^{*0} \rho^{0}$ & $12.3\pm0.4\pm0.5$
&$14.2\pm1.6\pm5$ & $13\pm2\pm2$\\
$D^{0} \rightarrow K^{-}a_{1}^{+}(1260)(\rho^{0}\pi^{+})$ & $54.6\pm2.8\pm3.7$
& $49.2\pm2.4\pm8$ & $47\pm5\pm10$\\
$D^{0} \rightarrow K_{1}^{-}(1270)(\bar{K}^{*0}\pi^{-})\pi^{+}$ & $0.8\pm0.2\pm0.2$
& \multirow{2}{*}{$6.6\pm1.9\pm3$} & \multirow{2}{*}{-} \\
$D^{0} \rightarrow K_{1}^{-}(1270)(K^{-}\rho^{0})\pi^{+}$  & $3.4\pm0.3\pm0.5$ & & \\
$D^{0} \rightarrow K^{-}\pi^{+}\rho^{0}$ & $8.4\pm1.1\pm2.5$
& $8.4\pm2.2\pm4$ & $5\pm3\pm2$\\
$D^{0} \rightarrow \bar{K}^{*0}\pi^{+}\pi^{-}$ & $7.0\pm0.4\pm0.5$
& $14.0\pm1.8\pm4$ & $11\pm2\pm3$\\
$D^{0} \rightarrow K^{-}\pi^{+}\pi^{+}\pi^{-}$ & $21.9\pm0.6\pm0.6$
& $24.2\pm2.5\pm6$ & $23\pm2\pm3$ \\
\hline
\end{tabular}
\label{tab:FF for comp}
\end{center}
\end{table*}

\begin{table*}[hbtp]
\footnotesize
\renewcommand\arraystretch{1.1}
\begin{center}
\caption{Phases and fit fractions for different amplitudes. 
The first and second uncertainties are statistical and systematic, respectively.}
\begin{tabular}{lcc} \hline
Amplitude & $\phi_{i}$ & Fit fraction (\%)\\ \hline
$D^{0}[S] \rightarrow \bar{K}^{*} \rho^{0}$&$ 2.35\pm0.06\pm0.18$&$6.5\pm0.5\pm0.8$\\
$D^{0}[P] \rightarrow \bar{K}^{*} \rho^{0}$&$-2.25\pm0.08\pm0.15$&$2.3\pm0.2\pm0.1$\\
$D^{0}[D] \rightarrow \bar{K}^{*} \rho^{0}$&$ 2.49\pm0.06\pm0.11$&$7.9\pm0.4\pm0.7$\\
$D^{0} \rightarrow K^{-}a_{1}^{+}(1260)$, $a_{1}^{+}(1260)[S] \rightarrow \rho^{0}\pi^{+}$ 
& $0$(fixed) & $53.2 \pm 2.8 \pm 4.0$ \\ 
$D^{0} \rightarrow K^{-}a_{1}^{+}(1260)$, $a_{1}^{+}(1260)[D] \rightarrow \rho^{0}\pi^{+}$ 
& $-2.11 \pm 0.15 \pm 0.21$ & $0.3 \pm 0.1 \pm 0.1$\\ 
$D^{0} \rightarrow K_{1}^{-}(1270)\pi^{+}$, $K_{1}^{-}(1270)[S] \rightarrow \bar{K}^{*0}\pi^{-}$ 
& $1.48 \pm 0.21 \pm 0.24$ & $0.1 \pm 0.1 \pm 0.1$ \\ 
$D^{0} \rightarrow K_{1}^{-}(1270)\pi^{+}$, $K_{1}^{-}(1270)[D] \rightarrow \bar{K}^{*0}\pi^{-}$ 
& $3.00 \pm 0.09 \pm 0.15$ & $0.7 \pm 0.2 \pm 0.2$ \\ 
$D^{0} \rightarrow K_{1}^{-}(1270)\pi^{+}$, $K_{1}^{-}(1270) \rightarrow K^{-}\rho^{0}$ 
& $-2.46 \pm 0.06 \pm 0.21$ & $3.4 \pm 0.3 \pm 0.5$ \\ 
$D^{0} \rightarrow (\rho^{0} K^{-})_{{\rm A}} \pi^{+}$, 
$(\rho^{0} K^{-})_{{\rm A}}[D] \rightarrow K^{-}\rho^{0}$  
& $-0.43 \pm 0.09 \pm 0.12$ & $1.1 \pm 0.2 \pm 0.3$ \\ 
$D^{0} \rightarrow (K^{-}\rho^{0})_{{\rm P}}\pi^{+}$ & 
$-0.14 \pm 0.11 \pm 0.10$ & $7.4 \pm 1.6 \pm 5.7$ \\ 
$D^{0} \rightarrow (K^{-}\pi^{+})_{{\rm S-wave}}\rho^{0}$ & 
$-2.45 \pm 0.19 \pm 0.47$ & $2.0 \pm 0.7 \pm 1.9$ \\ 
$D^{0} \rightarrow (K^{-}\rho^{0})_{{\rm V}}\pi^{+}$ & 
$-1.34 \pm 0.12 \pm 0.09$ & $0.4 \pm 0.1 \pm 0.1$ \\ 
$D^{0} \rightarrow (\bar{K}^{*0}\pi^{-})_{{\rm P}}\pi^{+}$ & 
$-2.09 \pm 0.12 \pm 0.22$ & $2.4 \pm 0.5 \pm 0.5$ \\ 
$D^{0} \rightarrow \bar{K}^{*0}(\pi^{+}\pi^{-})_{{\rm S}}$ & 
$-0.17 \pm 0.11 \pm 0.12$ & $2.6 \pm 0.6 \pm 0.6$ \\ 
$D^{0} \rightarrow (\bar{K}^{*0}\pi^{-})_{{\rm V}}\pi^{+} $ & 
$-2.13 \pm 0.10 \pm 0.11$ & $0.8 \pm 0.1 \pm 0.1$ \\ 
$D^{0} \rightarrow ((K^{-}\pi^{+})_{{\rm S-wave}}\pi^{-})_{{\rm A}}\pi^{+} $ & 
$-1.36 \pm 0.08 \pm 0.37$ & $5.6 \pm 0.9 \pm 2.7$ \\ 
$D^{0} \rightarrow K^{-}((\pi^{+}\pi^{-})_{{\rm S}}\pi^{+})_{{\rm A}} $ & 
$-2.23 \pm 0.08 \pm 0.22$ & $13.1 \pm 1.9 \pm 2.2$ \\ 
$D^{0} \rightarrow (K^{-}\pi^{+})_{{\rm S-wave}}(\pi^{+}\pi^{-})_{{\rm S}}$ & 
$-1.40 \pm 0.04 \pm 0.22$ & $16.3 \pm 0.5 \pm 0.6$ \\ 
$D^{0}[S] \rightarrow (K^{-}\pi^{+})_{{\rm V}}(\pi^{+}\pi^{-})_{{\rm V}} $ & 
$1.59 \pm 0.13 \pm 0.41$ & $5.4 \pm 1.2 \pm 1.9$ \\ 
$D^{0} \rightarrow (K^{-}\pi^{+})_{{\rm S-wave}}(\pi^{+}\pi^{-})_{{\rm V}}$ & 
$-0.16 \pm 0.17 \pm 0.43$ & $1.9 \pm 0.6 \pm 1.2$ \\ 
$D^{0} \rightarrow (K^{-}\pi^{+})_{{\rm V}}(\pi^{+}\pi^{-})_{{\rm S}}$ & 
$2.58 \pm 0.08 \pm 0.25$ & $2.9 \pm 0.5 \pm 1.7$ \\ 
$D^{0} \rightarrow (K^{-}\pi^{+})_{{\rm T}}(\pi^{+}\pi^{-})_{{\rm S}}$ & 
$-2.92 \pm 0.14 \pm 0.12$ &$0.3 \pm 0.1 \pm 0.1$ \\ 
$D^{0} \rightarrow (K^{-}\pi^{+})_{{\rm S-wave}}(\pi^{+}\pi^{-})_{{\rm T}}$ & 
$2.45 \pm 0.12 \pm 0.37$ & $0.5 \pm 0.1 \pm 0.1$ \\ 
\hline
\end{tabular}
\label{tab:Phase and FF}
\end{center}
\end{table*}

\section{Systematic Uncertainties}
\label{p_sys_unc}
The source of systematic uncertainties are divided into four categories: 
(I) amplitude model, (II) background estimation, 
(III) experimental effects and (IV) fitter performance.
The systematic uncertainties of the free parameters in the fit 
and the fit fractions due to different contributions
are given in units of the statistical standard deviations
$\sigma_{stat}$ in Tables~\ref{Tab: sys_res}--\ref{Tab: sys_FF2}.
These uncertainties are added in quadrature, as they are uncorrelated, 
to obtain the total systematic uncertainties.
\begin{table}[hbtp]
\footnotesize
\begin{center}
\caption{Systematic uncertainties on masses and widths of 
intermediate resonances $\bar{K}^{*0}$ and $\rho^{0}$.}
\begin{tabular}{cccccc} \hline
\multirow{2}{*}{Parameter} & \multicolumn{4}{c}{Source ($\sigma_{stat}$)} &\multirow{2}{*}{total ($\sigma_{stat}$)}\\
$~$ & I & II & III & IV & $~$\\ \hline
$m_{\bar{K}^{*0}}$&2.21&0.04&0.13&0.10&2.22\\ 
$\Gamma_{\bar{K}^{*0}}$&0.87&0.05&0.17&0.07&0.89\\ 
$m_{\rho^{0}}$&2.37&0.08&0.12&0.08&2.37\\ 
$\Gamma_{\rho^{0}}$&1.16&0.04&0.11&0.12&1.17\\ 
\hline
\end{tabular}
\label{Tab: sys_res}
\end{center}
\end{table}

\begin{table*}[hbtp]
\footnotesize
\begin{center}
\caption{Systematic uncertainties on fit fractions for different components.}
\begin{tabular}{lccccc} \hline
\multirow{2}{*}{Fit fraction} & \multicolumn{4}{c}{Source ($\sigma_{stat}$)} &\multirow{2}{*}{total ($\sigma_{stat}$)}\\
$~$ & I & II & III & IV & $~$\\ \hline
$D^{0} \rightarrow \bar{K}^{*0}\rho^{0}$&1.12&0.06&0.11&0.08&1.13\\
$D^{0} \rightarrow K^{-}a_{1}^{+}(1260)$&1.32&0.09&0.12&0.06&1.33\\
$D^{0} \rightarrow K_{1}^{-}(1270)(\bar{K}^{*0}\pi^{-})\pi^{+}$&1.41&0.02&0.12&0.10&1.42\\
$D^{0} \rightarrow K_{1}^{-}(1270)(K^{-}\rho^{0})\pi^{+}$&1.58&0.04&0.23&0.06&1.60\\
$D^{0} \rightarrow K^{-}\pi^{+}\rho^{0}$&2.22&0.10&0.12&0.15&2.23\\
$D^{0} \rightarrow \bar{K}^{*0}\pi^{+}\pi^{-}$&1.32&0.08&0.13&0.10&1.34\\
$D^{0} \rightarrow K^{-}\pi^{+}\pi^{+}\pi^{-}$&0.94&0.10&0.09&0.12&1.00\\
\hline
\end{tabular}
\label{tab: sys_FF}
\caption{Systematic uncertainties on phases and fit fractions for different amplitudes.}
\begin{tabular}{lccccc} \hline
\multirow{2}{*}{$\phi_{i}$} & \multicolumn{4}{c}{Source ($\sigma_{stat}$)} &\multirow{2}{*}{total ($\sigma_{stat}$)}\\
$~$ & I & II & III & IV & \\ \hline
$D^{0}[S] \rightarrow \bar{K}^{*0}\rho^{0}$&2.96&0.04&0.14&0.13&2.97\\
$D^{0}[P] \rightarrow \bar{K}^{*0}\rho^{0}$&1.98&0.04&0.11&0.12&1.98\\
$D^{0}[D] \rightarrow \bar{K}^{*0}\rho^{0}$&1.78&0.03&0.18&0.09&1.79\\
$D^{0} \rightarrow K^{-}a_{1}^{+}(1260)$, $a_{1}^{+}(1260)[D] \rightarrow \rho^{0}\pi^{+}$
&1.38&0.02&0.09&0.09&1.39\\ 
$D^{0} \rightarrow K_{1}^{-}(1270)\pi^{+}$,$K_{1}^{-}(1270)[S] \rightarrow \bar{K}^{*0}\pi^{-}$
&1.10&0.07&0.10&0.09&1.11\\ 
$D^{0} \rightarrow K_{1}^{-}(1270)\pi^{+}$,$K_{1}^{-}(1270)[D] \rightarrow \bar{K}^{*0}\pi^{-}$
&1.61&0.06&0.11&0.06&1.62\\ 
$D^{0} \rightarrow K_{1}^{-}(1270)\pi^{+}$,$K_{1}^{-}(1270) \rightarrow K^{-}\rho^{0}$
&3.61&0.03&0.09&0.13&3.62\\
$D^{0} \rightarrow (\rho^{0} K^{-})_{{\rm A}} \pi^{+}$&1.28&0.06&0.14&0.09&1.29\\ 
$D^{0} \rightarrow (K^{-}\rho^{0})_{{\rm P}}\pi^{+}$&0.92&0.10&0.10&0.07&0.93\\ 
$D^{0} \rightarrow (K^{-}\pi^{+})_{{\rm S-wave}}\rho^{0}$&2.46&0.06&0.10&0.09&2.47\\
$D^{0} \rightarrow (K^{-}\rho^{0})_{{\rm V}}\pi^{+}$&0.74&0.01&0.09&0.08&0.75\\
$D^{0} \rightarrow (\bar{K}^{*0}\pi^{-})_{{\rm P}}\pi^{+}$&1.82&0.03&0.09&0.06&1.82\\
$D^{0} \rightarrow \bar{K}^{*}(\pi^{+}\pi^{-})_{{\rm S}}$&1.07&0.04&0.12&0.11&1.08\\
$D^{0} \rightarrow (\bar{K}^{*0}\pi^{-})_{{\rm V}}\pi^{+}$&1.00&0.02&0.10&0.18&1.02\\
$D^{0} \rightarrow ((K^{-}\pi^{+})_{{\rm S-wave}}\pi^{-})_{{\rm A}}\pi^{+}$&4.78&0.15&0.12&0.07&4.79\\ 
$D^{0} \rightarrow K^{-}((\pi^{+}\pi^{-})_{{\rm S}}\pi^{+})_{{\rm A}}$&2.69&0.13&0.10&0.07&2.70\\ 
$D^{0} \rightarrow (K^{-}\pi^{+})_{{\rm S-wave}}(\pi^{+}\pi^{-})_{{\rm S}}$&6.27&0.04&0.10&0.12&6.27\\ 
$D^{0}[S] \rightarrow (K^{-}\pi^{+})_{{\rm V}}(\pi^{+}\pi^{-})_{{\rm V}}$&3.28&0.06&0.09&0.06&3.28\\
$D^{0} \rightarrow (K^{-}\pi^{+})_{{\rm S-wave}}(\pi^{+}\pi^{-})_{{\rm V}}$&2.59&0.09&0.10&0.10&2.60\\ 
$D^{0} \rightarrow (K^{-}\pi^{+})_{{\rm V}}(\pi^{+}\pi^{-})_{{\rm S}}$&3.07&0.09&0.10&0.18&3.08\\ 
$D^{0} \rightarrow (K^{-}\pi^{+})_{{\rm T}}(\pi^{+}\pi^{-})_{{\rm S}}$&0.81&0.04&0.12&0.06&0.82\\ 
$D^{0} \rightarrow (K^{-}\pi^{+})_{{\rm S-wave}}(\pi^{+}\pi^{-})_{{\rm T}}$&3.11&0.06&0.11&0.16&3.19\\ 
\hline
\end{tabular}
\begin{tabular}{lccccc} 
\multirow{2}{*}{Fit fraction} & \multicolumn{4}{c}{Source ($\sigma_{stat}$)} &\multirow{2}{*}{total ($\sigma_{stat}$)}\\
$~$ & I & II & III & IV\\ \hline
$D^{0}[S] \rightarrow \bar{K}^{*0}\rho^{0}$&1.76&0.04&0.09&0.10&1.77\\
$D^{0}[P] \rightarrow \bar{K}^{*0}\rho^{0}$&0.27&0.02&0.09&0.12&0.31\\
$D^{0}[D] \rightarrow \bar{K}^{*0}\rho^{0}$&1.79&0.06&0.12&0.17&1.80\\
$D^{0} \rightarrow K^{-}a_{1}^{+}(1260)$, $a_{1}^{+}(1260)[S] \rightarrow \rho^{0}\pi^{+}$
&1.48&0.10&0.12&0.07&1.45\\
$D^{0} \rightarrow K^{-}a_{1}^{+}(1260)$, $a_{1}^{+}(1260)[D] \rightarrow \rho^{0}\pi^{+}$
&0.93&0.04&0.09&0.06&0.94\\
$D^{0} \rightarrow K_{1}^{-}(1270)\pi^{+}$,$K_{1}^{-}(1270)[S] \rightarrow \bar{K}^{*0}\pi^{-}$
&1.01&0.05&0.11&0.16&1.03\\
$D^{0} \rightarrow K_{1}^{-}(1270)\pi^{+}$,$K_{1}^{-}(1270)[D] \rightarrow \bar{K}^{*0}\pi^{-}$
&1.12&0.03&0.12&0.13&1.14\\
$D^{0} \rightarrow K_{1}(1270)^{-}\pi^{+}$,$K_{1}^{-}(1270) \rightarrow K^{-}\rho^{0}$
&1.58&0.04&0.23&0.06&1.60\\
$D^{0} \rightarrow (\rho^{0} K^{-})_{{\rm A}} \pi^{+}$&1.38&0.08&0.09&0.09&1.39\\
$D^{0} \rightarrow (\bar{K}^{*0}\pi)_{{\rm P}}\pi$&0.93&0.06&0.09&0.16&0.95\\
$D^{0} \rightarrow (K^{-}\pi^{+})_{{\rm S-wave}}\rho^{0}$&2.81&0.09&0.11&0.09&2.82\\
$D^{0} \rightarrow (K^{-}\rho^{0})_{{\rm V}}\pi^{+}$&0.69&0.03&0.09&0.06&0.70\\
$D^{0} \rightarrow (\bar{K}^{*0}\pi^{-})_{{\rm P}}\pi^{+}$&0.93&0.06&0.09&0.16&0.95\\
$D^{0} \rightarrow \bar{K}^{*0}(\pi^{+}\pi^{-})_{{\rm S}}$&1.06&0.05&0.09&0.20&1.08\\
$D^{0} \rightarrow (\bar{K}^{*0}\pi^{-})_{{\rm V}}\pi^{+}$&0.60&0.02&0.00&0.10&0.61\\
$D^{0} \rightarrow ((K^{-}\pi^{+})_{{\rm S-wave}}\pi^{-})_{{\rm A}}\pi^{+}$&3.10&0.07&0.09&0.06&3.10\\
$D^{0} \rightarrow K^{-}((\pi^{+}\pi^{-})_{{\rm S}}\pi^{+})_{{\rm A}}$&1.14&0.08&0.10&0.07&1.15\\
$D^{0} \rightarrow (K^{-}\pi^{+})_{{\rm S-wave}}(\pi^{+}\pi^{-})_{{\rm S}}$&1.29&0.12&0.10&0.12&1.30\\
$D^{0}[S] \rightarrow (K^{-}\pi^{+})_{{\rm V}}(\pi^{+}\pi^{-})_{{\rm V}}$&1.73&0.07&0.09&0.07&1.73\\
$D^{0} \rightarrow (K^{-}\pi^{+})_{{\rm S-wave}}(\pi^{+}\pi^{-})_{{\rm V}}$&2.08&0.12&0.10&0.07&2.09\\
$D^{0} \rightarrow (K^{-}\pi^{+})_{{\rm V}}(\pi^{+}\pi^{-})_{{\rm S}}$&3.54&0.05&0.10&0.11&3.54\\
$D^{0} \rightarrow (K^{-}\pi^{+})_{{\rm T}}(\pi^{+}\pi^{-})_{{\rm S}}$&0.87&0.07&0.11&0.07&0.88\\
$D^{0} \rightarrow (K^{-}\pi^{+})_{{\rm S-wave}}(\pi^{+}\pi^{-})_{{\rm T}}$&0.99&0.09&0.10&0.08&1.01\\
\hline
\end{tabular}
\label{Tab: sys_FF2}
\end{center}
\end{table*}
\subsubsection{Amplitude model}
\label{Model_sys}
Three sources are considered for the systematic uncertainty due to the amplitude model: the 
masses and widths of the $K_{1}^{-}(1270)$ and the $a_{1}^{+}(1260)$, the barrier effective radius $R$ 
and the fixed parameters in the $K\pi$ $S$-wave model. 
The uncertainty associated with the mass and width of $K_{1}^{-}(1270)$ and the $a_{1}^{+}(1260)$ are 
estimated by varying the corresponding masses and widths 
with 1$\sigma$ of errors quoted in PDG~\cite{PDG}, respectively.
The uncertainty related to the barrier effective radius $R$ is estimated by varying $R$ within 
$1.5-4.5 {\mbox{\,GeV}}^{-1}$ for the intermediate resonances and 
$3.0-7.0{\mbox{\,GeV}}^{-1}$ for the $D^{0}$ in the fit. 
The uncertainty from the input parameters of the $K\pi$ $S$-wave model are evaluated by varying the input values 
within their uncertainties.
All the change of the results with respect to the nominal one are taken as the systematic uncertainties.

\subsubsection{Background estimation}
\label{BKG_sys}
The sources of systematic uncertainty related to the background include the amplitude and shape of 
the background $D^{0} \rightarrow K_{S}^{0}K^{-}\pi^{+}$, and the
other potential backgrounds.
The uncertainties related to the background $D^{0} \rightarrow K_{S}^{0}K^{-}\pi^{+}$ is 
estimated by varying the number of background events within 1$\sigma$ of uncertainties and changing the shape 
according to the uncertainties in PDF parameters from CLEO~\cite{KsKPi}. 
The uncertainty due to the the other potential background is estimated by including the corresponding 
background (estimated from generic MC sample) in the fit.

\subsubsection{Experimental effects}
The uncertainty related to the experimental effects includes two separate components: 
the acceptance difference between MC simulations and data caused by tracking and PID efficiencies, 
and the detector resolution.
To determine the systematic uncertainty due to tracking and PID efficiencies, 
we alter the fit by shifting the $\gamma_{\epsilon}(p)$ in Eq.~(\ref{trackingPID})  
within its uncertainty, and the changes of the nominal results 
is taken as the systematic uncertainty.
The uncertainty caused by resolution is determined as the difference 
between the pull distribution results obtained from simulated data using 
generated and fitted four-momenta, as described in Sec.~\ref{Fitter Performance}. 

\subsubsection{Fitter performance}
\label{Fitter Performance}
The uncertainty from the fit process is evaluated by studying toy MC samples. 
An ensemble of 250 sets of SIGNAL MC samples with a size equal to the data sample are generated according to the 
nominal results in this analysis.
The SIGNAL MC samples are fed into the event selection, and the same amplitude analysis is performed
on each simulated sample. The pull variables, $\frac{V_{{\rm input}}-V_{{\rm fit}}}{\sigma_{{\rm fit}}}$, are 
defined to evaluate the corresponding uncertainty, where $V_{{\rm input}}$ is the input value 
in the generator, $V_{{\rm fit}}$ and $\sigma_{{\rm fit}}$ are the output value and the corresponding 
statistical uncertainty, respectively. 
The distribution of pull values for the 250 sets of sample are expected to be a normal Gaussian distribution, 
and any shift on mean and widths indicate the bias on the fit values and its 
statistical uncertainty, respectively.

Small biases for some fitted parameters 
and fit fractions are observed. For the pull mean,
the largest bias is about 19\% of a statistical uncertainty with a
deviation of about 3.0$\sigma$ from zero. For the pull width, the largest shift 
is $0.87\pm0.04$, about 3.0 standard deviations from 1.0.  
We add in quadrature the mean and the mean error 
in the pull and multiply this number with the statistical error to get the systematic error. 
The fit results are given in Tables~\ref{Pull: res}$\sim$\ref{Pull: FF2}. 
The uncertainties in Tables~\ref{Pull: res}$\sim$\ref{Pull: FF2} are the statistical uncertainties of the 
fits to the pull distributions.
\begin{table*}[hbtp]
\footnotesize
\begin{center}
\caption{Pull mean and pull width of the pull distributions for the fitted masses and 
widths of intermediate resonances $\bar{K}^{*0}$ and $\rho^{0}$ from
simulated data using either the generated or fitted four-momenta.}
\begin{tabular}{ccccc} \hline
\multirow{2}{*}{Parameter} & \multicolumn{2}{c}{Generated $p_{i}$} & \multicolumn{2}{c}{Fitted $p_{i}$} \\
$~$ & pull mean & pull width & pull mean & pull width \\ \hline
$m_{\bar{K}^{*0}}$ & $0.07\pm0.07$ & $1.05\pm0.05$ & $0.06\pm0.07$ & $1.04\pm0.05$ \\
$\Gamma_{\bar{K}^{*0}}$ & $-0.03\pm0.06$ & $0.97\pm0.04$ & $-0.17\pm0.06$&$0.97\pm0.04$\\
$m_{\rho^{0}}$ & $0.03\pm0.07$ & $1.06\pm0.05$ & $-0.02\pm0.07$ & $1.06\pm0.05$\\
$\Gamma_{\rho^{0}}$ & $0.10\pm0.07$ & $1.08\pm0.05$ & $0.06\pm0.07$ & $1.07\pm0.05$\\
\hline
\end{tabular}
\label{Pull: res}
\caption{Pull mean and pull width of the pull distributions for the different components from
simulated data using either the generated or fitted four-momenta.}
\begin{tabular}{lcccc} \hline
\multirow{2}{*}{Fit fraction} & \multicolumn{2}{c}{Generated $p_{i}$} & \multicolumn{2}{c}{Fitted $p_{i}$} \\
$~$ & pull mean & pull width & pull mean & pull width \\ \hline
$D^{0} \rightarrow \bar{K}^{*0}\rho^{0}$&$0.05\pm0.06$&$0.92\pm0.04$&$0.04\pm0.06$&$0.89\pm0.04$\\
$D^{0} \rightarrow K^{-}a_{1}^{+}(1260)$&$0.02\pm0.06$&$0.91\pm0.04$&$0.04\pm0.06$&$0.87\pm0.04$\\
$D^{0} \rightarrow K_{1}^{-}(1270)(\bar{K}^{*0}\pi^{-})\pi^{+}$ 
&$-0.08\pm0.06$&$0.98\pm0.04$&$-0.06\pm0.06$&$0.97\pm0.04$\\
$D^{0} \rightarrow K_{1}^{-}(1270)(K^{-}\rho^{0})\pi^{+}$
&$0.01\pm0.06$&$0.98\pm0.04$&$0.01\pm0.06$&$0.99\pm0.04$\\
$D^{0} \rightarrow K^{-}\pi^{+}\rho^{0}$&$0.14\pm0.06$&$0.92\pm0.04$&$0.11\pm0.06$&$0.88\pm0.04$\\
$D^{0} \rightarrow \bar{K}^{*0}\pi^{+}\pi^{-}$&$-0.08\pm0.06$&$0.96\pm0.04$&$-0.09\pm0.06$&$0.96\pm0.04$\\
$D^{0} \rightarrow K^{-}\pi^{+}\pi^{+}\pi^{-}$&$0.10\pm0.06$&$0.94\pm0.04$&$0.12\pm0.06$&$0.93\pm0.04$\\
\hline
\end{tabular}
\label{Pull: FF}
\caption{Pull mean and pull width of the pull distributions 
for the phases and fit fractions of different amplitudes, 
from simulated data using either the generated or fitted four-momenta.}
\begin{tabular}{lcccc} \hline
\multirow{2}{*}{$\phi_{i}$} & \multicolumn{2}{c}{Generated $p_{i}$} & \multicolumn{2}{c}{Fitted $p_{i}$} \\
$~$ & pull mean & pull width & pull mean & pull width \\ \hline
$D^{0}[S] \rightarrow \bar{K}^{*0}\rho^{0}$&$0.11\pm0.06$&$1.01\pm0.05$&$0.08\pm0.06$&$1.00\pm0.04$\\
$D^{0}[P] \rightarrow \bar{K}^{*0}\rho^{0}$&$0.10\pm0.07$&$1.03\pm0.05$&$0.08\pm0.06$&$1.02\pm0.05$\\
$D^{0}[D] \rightarrow \bar{K}^{*0}\rho^{0}$&$0.05\pm0.07$&$1.04\pm0.05$&$0.01\pm0.07$&$1.03\pm0.05$\\
$D^{0} \rightarrow K^{-}a_{1}^{+}(1260)$, $a_{1}^{+}(1260)[D] \rightarrow \rho^{0}\pi^{+}$
& $-0.07\pm0.06$ & $1.02\pm0.05$ & $-0.05\pm0.06$ & $1.02\pm0.05$\\
$D^{0} \rightarrow K_{1}^{-}(1270)\pi^{+}$,$K_{1}^{-}(1270)[S] \rightarrow \bar{K}^{*0}\pi^{-}$
& $0.06\pm0.07$ & $1.03\pm0.05$ & $0.06\pm0.06$ & $1.03\pm0.05$\\
$D^{0} \rightarrow K_{1}^{-}(1270)\pi^{+}$,$K_{1}^{-}(1270)[D] \rightarrow \bar{K}^{*0}\pi^{-}$
&$-0.02\pm0.06$&$0.98\pm0.04$ & $-0.06\pm0.06$ & $0.97\pm0.04$\\
$D^{0} \rightarrow K_{1}^{-}(1270)\pi^{+}$,$K_{1}^{-}(1270) \rightarrow K^{-}\rho^{0}$
&$0.12\pm0.06$&$1.00\pm0.04$&$0.11\pm0.06$&$1.00\pm0.04$\\
$D^{0} \rightarrow (\rho^{0} K^{-})_{{\rm A}} \pi^{+}$&$-0.06\pm0.07$&$1.05\pm0.05$&$-0.09\pm0.07$&$1.05\pm0.05$\\
$D^{0} \rightarrow (K^{-}\rho^{0})_{{\rm P}}\pi^{+}$&$-0.03\pm0.06$&$0.96\pm0.04$&$-0.01\pm0.06$&$0.96\pm0.04$\\
$D^{0} \rightarrow (K^{-}\pi^{+})_{{\rm S-wave}}\rho^{0}$&$-0.07\pm0.06$&$0.92\pm0.04$&$-0.08\pm0.06$&$0.92\pm0.04$\\
$D^{0} \rightarrow (K^{-}\rho^{0})_{{\rm V}}\pi^{+}$&$-0.05\pm0.06$&$1.02\pm0.05$&$-0.07\pm0.06$&$1.01\pm0.05$\\
$D^{0} \rightarrow (\bar{K}^{*0}\pi^{-})_{{\rm P}}\pi^{+}$&$0.00\pm0.06$&$0.99\pm0.04$&$0.00\pm0.06$&$0.99\pm0.04$\\
$D^{0} \rightarrow \bar{K}^{*0}(\pi^{+}\pi^{-})_{{\rm S}}$&$-0.08\pm0.07$&$1.03\pm0.05$&$-0.11\pm0.07$&$1.03\pm0.05$\\
$D^{0} \rightarrow (\bar{K}^{*0}\pi^{-})_{{\rm V}}\pi^{+}$&$0.17\pm0.06$&$0.99\pm0.04$&$0.15\pm0.06$&$0.98\pm0.04$\\
$D^{0} \rightarrow ((K^{-}\pi^{+})_{{\rm S-wave}}\pi^{-})_{{\rm A}}\pi^{+}$&
$-0.04\pm0.06$&$0.92\pm0.04$&$0.02\pm0.06$&$0.92\pm0.04$\\
$D^{0} \rightarrow K^{-}((\pi^{+}\pi^{-})_{{\rm S}}\pi^{+})_{{\rm A}}$&
$0.00\pm0.07$&$1.05\pm0.05$&$-0.02\pm0.07$&$1.04\pm0.05$\\
$D^{0} \rightarrow (K^{-}\pi^{+})_{{\rm S-wave}}(\pi^{+}\pi^{-})_{{\rm S}}$&
$0.10\pm0.06$&$0.98\pm0.04$&$0.08\pm0.06$&$0.98\pm0.04$\\
$D^{0}[S] \rightarrow (K^{-}\pi^{+})_{{\rm V}}(\pi^{+}\pi^{-})_{{\rm V}}$&
$-0.02\pm0.06$&$0.97\pm0.04$&$-0.03\pm0.06$&$0.98\pm0.04$\\
$D^{0} \rightarrow (K^{-}\pi^{+})_{{\rm S-wave}}(\pi^{+}\pi^{-})_{{\rm V}}$&$0.08\pm0.06$&$0.93\pm0.04$&$0.06\pm0.06$&$0.92\pm0.04$\\
$D^{0} \rightarrow (K^{-}\pi^{+})_{{\rm V}}(\pi^{+}\pi^{-})_{{\rm S}}$&$-0.17\pm0.06$&$0.94\pm0.04$&$-0.17\pm0.06$&$0.94\pm0.04$\\
$D^{0} \rightarrow (K^{-}\pi^{+})_{{\rm T}}(\pi^{+}\pi^{-})_{{\rm S}}$&$0.01\pm0.06$&$1.01\pm0.05$&$-0.02\pm0.06$&$1.00\pm0.04$\\
$D^{0} \rightarrow (K^{-}\pi^{+})_{{\rm S-wave}}(\pi^{+}\pi^{-})_{{\rm T}}$&$0.14\pm0.07$&$1.12\pm0.05$&$0.12\pm0.07$&$1.11\pm0.05$\\
\hline
\end{tabular}
\begin{tabular}{lcccc} 
\multirow{2}{*}{Fit fraction}& \multicolumn{2}{c}{Generated $p_{i}$} & \multicolumn{2}{c}{Fitted $p_{i}$} \\
$~$ & pull mean & pull width & pull mean & pull width \\ \hline
$D^{0}[S] \rightarrow \bar{K}^{*0}\rho^{0}$&$0.08\pm0.06$&$0.88\pm0.04$&$0.07\pm0.06$&$0.87\pm0.04$\\
$D^{0}[P] \rightarrow \bar{K}^{*0}\rho^{0}$&$0.10\pm0.06$&$0.97\pm0.04$&$0.10\pm0.06$&$0.96\pm0.04$\\
$D^{0}[D] \rightarrow \bar{K}^{*0}\rho^{0}$&$-0.15\pm0.07$&$1.10\pm0.05$&$-0.15\pm0.07$&$1.10\pm0.05$\\
$D^{0} \rightarrow K^{-}a_{1}^{+}(1260)$, $a_{1}^{+}(1260)[S] \rightarrow \rho^{0}\pi^{+}$
&$0.03\pm0.06$&$0.91\pm0.04$&$0.04\pm0.06$&$0.90\pm0.04$\\
$D^{0} \rightarrow K^{-}a_{1}^{+}(1260)$, $a_{1}^{+}(1260)[D] \rightarrow \rho^{0}\pi^{+}$
&$0.02\pm0.06$&$1.00\pm0.04$&$0.03\pm0.06$&$1.00\pm0.04$\\
$D^{0} \rightarrow K_{1}^{-}(1270)\pi$,$K_{1}^{-}(1270)[S] \rightarrow \bar{K}^{*0}\pi^{-}$
&$-0.14\pm0.07$&$1.02\pm0.05$&$-0.18\pm0.07$&$1.09\pm0.05$\\
$D \rightarrow K_{1}^{-}(1270)\pi^{+}$,$K_{1}^{-}(1270)[D] \rightarrow \bar{K}^{*0}\pi^{-}$
&$-0.11\pm0.06$&$0.99\pm0.04$&$-0.09\pm0.06$&$0.99\pm0.04$\\
$D^{0} \rightarrow K_{1}^{-}(1270)\pi^{+}$,$K_{1}^{-}(1270) \rightarrow K^{-}\rho^{0}$
&$0.01\pm0.06$&$0.98\pm0.04$&$0.01\pm0.06$&$0.98\pm0.04$\\
$D^{0} \rightarrow (\rho^{0} K^{-})_{{\rm A}} \pi^{+}$&$0.06\pm0.06$&$1.00\pm0.04$&$0.04\pm0.06$&$0.99\pm0.04$\\
$D^{0} \rightarrow (K^{-}\rho^{0})_{{\rm P}}\pi^{+}$&$0.11\pm0.06$&$0.95\pm0.04$&$0.09\pm0.06$&$0.94\pm0.04$\\
$D^{0} \rightarrow (K^{-}\pi^{+})_{{\rm S-wave}}\rho^{0}$&$0.05\pm0.07$&$1.04\pm0.05$&$0.05\pm0.07$&$1.04\pm0.05$\\
$D^{0} \rightarrow (K^{-}\rho^{0})_{{\rm V}}\pi^{+}$&$0.01\pm0.06$&$0.98\pm0.04$&$0.02\pm0.06$&$0.97\pm0.04$\\
$D^{0} \rightarrow (\bar{K}^{*0}\pi^{-})_{{\rm P}}\pi^{+}$&$0.15\pm0.06$&$0.93\pm0.04$&$0.15\pm0.06$&$0.93\pm0.04$\\
$D^{0} \rightarrow \bar{K}^{*0}(\pi^{+}\pi^{-})_{{\rm S}}$&$-0.19\pm0.06$&$1.03\pm0.05$&$-0.18\pm0.06$&$1.02\pm0.05$\\
$D^{0} \rightarrow (\bar{K}^{*0}\pi^{-})_{{\rm V}}\pi^{+}$&$-0.08\pm0.06$&$1.00\pm0.04$&$-0.09\pm0.06$&$1.00\pm0.04$\\
$D^{0} \rightarrow ((K^{-}\pi^{+})_{{\rm S-wave}}\pi^{-})_{{\rm A}}\pi^{+}$
&$0.02\pm0.06$&$0.98\pm0.04$&$0.02\pm0.06$&$0.97\pm0.04$\\
$D^{0} \rightarrow K^{-}((\pi^{+}\pi^{-})_{{\rm S}}\pi^{+})_{{\rm A}}$
&$0.04\pm0.06$&$1.01\pm0.05$&$0.04\pm0.06$&$1.00\pm0.04$\\
$D^{0} \rightarrow (K^{-}\pi^{+})_{{\rm S-wave}}(\pi^{+}\pi^{-})_{{\rm S}}$
&$-0.10\pm0.06$&$0.93\pm0.04$&$-0.09\pm0.06$&$0.93\pm0.04$\\
$D^{0}[S] \rightarrow (K^{-}\pi^{+})_{{\rm V}}(\pi^{+}\pi^{-})_{{\rm V}}$
&$0.03\pm0.06$&$1.02\pm0.05$&$0.03\pm0.06$&$1.01\pm0.05$\\
$D^{0} \rightarrow (K^{-}\pi^{+})_{{\rm S-wave}}(\pi^{+}\pi^{-})_{{\rm V}}$&$0.04\pm0.06$&$1.00\pm0.04$&$0.04\pm0.06$&$0.99\pm0.04$\\
$D^{0} \rightarrow (K^{-}\pi^{+})_{{\rm V}}(\pi^{+}\pi^{-})_{{\rm S}}$&$0.09\pm0.07$&$1.06\pm0.05$&$0.11\pm0.07$&$1.04\pm0.05$\\
$D^{0} \rightarrow (K^{-}\pi^{+})_{{\rm T}}(\pi^{+}\pi^{-})_{{\rm S}}$&$0.01\pm0.07$&$1.05\pm0.05$&$0.00\pm0.07$&$1.03\pm0.05$\\
$D^{0} \rightarrow (K^{-}\pi^{+})_{{\rm S-wave}}(\pi^{+}\pi^{-})_{{\rm T}}$&$0.05\pm0.06$&$0.96\pm0.04$&$0.05\pm0.06$&$0.96\pm0.04$\\
\hline
\end{tabular}
\label{Pull: FF2}
\end{center}
\end{table*}

\section{Conclusion}
\label{CONLUSION}
An amplitude analysis of the decay $D^{0}\rightarrow K^{-}\pi^{+}\pi^{+}\pi^{-}$ has been performed with the 2.93 
${\mbox{\,fb}^{-1}}$ of $e^{+}e^{-}$ collision data at the $\psi(3770)$ resonance collected by the BESIII detector. 
The dominant components,
$D^{0}\rightarrow K^{-}a_{1}^{+}(1260)$, $D^{0}\rightarrow \bar{K}^{*0}\rho^{0}$, $D^{0}\rightarrow$ 
four-body nonresonant decay and three-body nonresonant $D^{0}\rightarrow K^{-}\pi^{+}\rho^{0}$ 
improve upon the earlier results from Mark III and are 
consistent with them within corresponding uncertainties. 
The resonance $K_{1}^{-}(1270)$ observed 
by Mark III is also confirmed in this analysis. 
The detailed results are listed in Table~\ref{tab:FF for comp}.  

About 40\% of components comes from the nonresonant four-body ($D^{0}\rightarrow K^{-}\pi^{+}\pi^{+}\pi^{-}$) 
and three-body ($D^{0} \rightarrow K^{-}\pi^{+}\rho^{0}$ and $D^{0} \rightarrow \bar{K}^{*0}\pi^{+}\pi^{-}$) 
decays. A detailed study considering the different orbital angular momentum is performed, 
which was not included in the analyses of Mark III and E691. An
especially interesting process involving the 
$K\pi$ S-wave is described by an effective range parametrization.

By using the inclusive branching fraction 
$\mathcal{B}(D^{0}\rightarrow K^{-}\pi^{+}\pi^{+}\pi^{-}) = (8.07 \pm 0.23)\%$ taken from the PDG~\cite{PDG} 
and the fit fraction for the different components $FF(n)$ obtained in this analysis,
we calculate the exclusive absolute branching fractions for the individual components with 
$\mathcal{B}(n) = \mathcal{B}(D^{0}\rightarrow K^{-}\pi^{+}\pi^{+}\pi^{-}) \times FF(n)$. 
The results are summarized in Table~\ref{tab:final res_BR} and are compared with the values quoted in PDG. 
Our results have much improved precision; they may shed light in a 
theoretical calculation. 
The knowledge of $D^{0} \rightarrow \bar{K}^{*0} \rho^{0}$ 
and $D^{0} \rightarrow K^{-} a_{1}^{+}(1260)$
increase our understanding of the decay $D^{0} \rightarrow VV$ and $D \rightarrow AP$, 
both of which are lacking in experimental measurements, but have large contributions to the $D^{0}$ decays. 
Furthermore, knowledge of the submodes in the decay $D^{0}\rightarrow K^{-}\pi^{+}\pi^{+}\pi^{-}$ 
will improve the determination of the reconstruction efficiency 
when this mode is used to tag the $D^{0}$ as part of other measurements, like measurements of 
branching fractions, the strong phase or the angle $\gamma$.

\begin{table}[htbp]
\scriptsize 
\begin{center}
\caption{Absolute branching fractions of the seven components and the 
corresponding values in the PDG. Here, we denote $\bar{K}^{*0}\rightarrow K^{-}\pi^{+}$ and 
$\rho^{0} \rightarrow \pi^{+}\pi^{-}$. 
The first two uncertainties are statistical and systematic, respectively. 
The third uncertainties are propagated from the uncertainty of  
$\mathcal{B}(D^{0} \rightarrow K^{-}\pi^{+}\pi^{+}\pi^{-})$. 
}
\begin{tabular}{llcc} \hline
Component & Branching fraction (\%) & PDG value (\%)\\ \hline
$D^{0} \rightarrow \bar{K}^{*0} \rho^{0}$ 
& $0.99\pm0.04\pm0.04\pm0.03$ & $1.05\pm0.23$ \\  
$D^{0} \rightarrow K^{-}a_{1}^{+}(1260)(\rho^{0}\pi^{+})$ 
& $4.41\pm0.22\pm0.30\pm0.13$ & $3.6\pm0.6$\\  
$D^{0} \rightarrow K_{1}^{-}(1270)(\bar{K}^{*0}\pi^{-})\pi^{+}$ 
& $0.07\pm0.01\pm0.02\pm0.00$ & \multirow{2}{*}{$0.29\pm0.03$} \\ 
$D^{0} \rightarrow K_{1}^{-}(1270)(K^{-}\rho^{0})\pi^{+}$
& $0.27\pm0.02\pm0.04\pm0.01$ & $~$ \\ 
$D^{0} \rightarrow K^{-}\pi^{+}\rho^{0}$ 
& $0.68\pm0.09\pm0.20\pm0.02$ & $0.51\pm0.23$ \\ 
$D^{0} \rightarrow \bar{K}^{*0}\pi^{+}\pi^{-}$ 
& $0.57\pm0.03\pm0.04\pm0.02$ & $0.99\pm0.23$ \\ 
$D^{0} \rightarrow K^{-}\pi^{+}\pi^{+}\pi^{-}$ 
& $1.77\pm0.05\pm0.04\pm0.05$ & $1.88\pm0.26$ \\ 
\hline
\end{tabular}
\label{tab:final res_BR}
\end{center}
\end{table}

\begin{acknowledgements}
\label{sec:acknowledgement}
\vspace{-0.4cm}

The BESIII collaboration thanks the staff of BEPCII
and the IHEP computing center for their strong support.
This work is supported in part by National Key Basic Research Program
of China under Contract No.~2015CB856700;
National Natural Science Foundation of China (NSFC) under Contracts
No.~11075174, No.~11121092, No.~11125525, No.~11235011, No.~11322544, No.~11335008, No.~11375221, No.~11425524, No.~11475185, No.~11635010;
the Chinese Academy of Sciences (CAS) Large-Scale Scientific Facility Program;
Joint Large-Scale Scientific Facility Funds of the NSFC and CAS under Contracts
No.~11179007, No.~U1232201, No.~U1332201;
CAS under Contracts No.~KJCX2-YW-N29, No.~KJCX2-YW-N45;
100 Talents Program of CAS;
INPAC and Shanghai Key Laboratory for Particle Physics and Cosmology;
German Research Foundation DFG under Contract No. Collaborative Research Center CRC-1044;
Istituto Nazionale di Fisica Nucleare, Italy;
Ministry of Development of Turkey under Contract No.~DPT2006K-120470;
Russian Foundation for Basic Research under Contract No.~14-07-91152;
U. S. Department of Energy under Contracts
No.~DE-FG02-04ER41291, No.~DE-FG02-05ER41374, No.~DE-FG02-94ER40823, No.~DESC0010118;
U.S. National Science Foundation;
University of Groningen (RuG) and the Helmholtzzentrum fuer Schwerionenforschung GmbH (GSI), Darmstadt;
WCU Program of National Research Foundation of Korea under Contract No.~R32-2008-000-10155-0.

\end{acknowledgements}

\section{Appendix A: Amplitudes Tested}
\label{sec:appenA}
The amplitudes listed below are tested when determining the nominal fit model, 
but not used in our final fit result.\\ 
{\bf Cascade amplitudes}\\
$K_{1}^{-}(1270)(\rho^{0} K^{-})\pi^{+}$, $\rho^{0} K^{-}$ $D$-wave\\
$K_{1}^{-}(1400)(\bar{K}^{*0}\pi^{-})\pi^{+}$, $\bar{K}^{*0}\pi^{-}$ $S$ and $D$-waves\\
$K^{*-}(1410)(\bar{K}^{*0}\pi^{-})\pi^{+}$\\
$K^{*-}_{2}(1430)(\bar{K}^{*0}\pi^{-})\pi^{+}$, $K^{*-}_{2}(1430)(K^{-}\rho^{0})\pi^{+}$\\
$K^{*-}(1680)(\bar{K}^{*0}\pi^{-})\pi^{+}$, $K^{*-}(1680)(K^{-}\rho^{0})\pi^{+}$\\
$K^{*-}_{2}(1770)(\bar{K}^{*0}\pi^{-})\pi^{+}$, $K^{*-}_{2}(1770)(K^{-}\rho^{0})\pi^{+}$\\
$K^{-}a_{2}^{+}(1320)(\rho^{0}\pi^{+})$\\
$K^{-}\pi^{+}(1300)(\rho^{0}\pi^{+})$\\
$K^{-}a_{1}^{+}(1260)(f_{0}(500)\pi^{+})$\\
{\bf Quasi-two-body amplitudes}\\
$\bar{K}^{*0}f_{0}(500)$\\
$\bar{K}^{*0}f_{0}(980)$\\
{\bf Three-body amplitudes}\\
$\bar{K}^{*0}(\pi^{+}\pi^{-})_{{\rm V}}$ $S$, $P$- and $D$-waves\\
$(K^{-}\pi^{+})_{{\rm V}}\rho^{0}$ $S$, $P$ and $D$-waves\\
$\bar{K}^{*0}_{2}(1430)(\pi^{+}\pi^{-})_{{\rm S}}$\\
$\bar{K}^{*0}_{2}(1430)\rho^{0}$\\
$\bar{K}^{*0}f_{2}(1270)$\\
$(K^{-}\pi^{+})_{{\rm S}}f_{2}(1270)$\\
$K^{-}(\rho^{0}\pi^{+})_{{\rm V}}$\\
$K^{-}(\rho^{0}\pi^{+})_{{\rm P}}$\\
$K^{-}(\rho^{0}\pi^{+})_{{\rm A}}$\\
$K^{-}(\rho^{0}\pi^{+})_{{\rm T}}$\\
$(\bar{K}^{*0}\pi^{-})_{{\rm T}}\pi^{+}$\\
$(K^{-}\rho^{0})_{{\rm T}}\pi^{+}$\\
$(\bar{K}^{*0}\pi^{-})_{{\rm A}}\pi^{+}$, $\bar{K}^{*0}\pi^{-}$ $S$ and $D$-waves\\
{\bf Four-body nonresonance amplitudes}\\
$(K^{-}\pi^{+})_{{\rm T}}(\pi^{+}\pi^{-})_{{\rm V}}$ $P$- and $D$-waves\\
$(K^{-}\pi^{+})_{{\rm V}}(\pi^{+}\pi^{-})_{{\rm T}}$ $P$- and $D$-waves\\
$(K^{-}\pi^{+})_{{\rm V}}(\pi^{+}\pi^{-})_{{\rm V}}$ $P$- and $D$-waves\\
$(K^{-}(\pi^{+}\pi^{-})_{{\rm S}})_{{\rm A}}\pi^{+}$\\


\begin{thebibliography}{99}

\bibitem{PDG} C. Patrignani {\it et al.} (Particle Data Group), Chin. Phys. C {\bf 40}, 100001 (2016).

\bibitem{CLEODdecay} G. Bonvicini {\it et al}. (CLEO Collaboration), Phys. Rev. D {\bf 89}, 072002 (2014).

\bibitem{ADS} D. Atwood, I. Dunietz and A. Soni, Phys. Rev. Lett. {\bf 78}, 3257 (1997).

\bibitem{K3Pigamma} S. Harnew and J. Rademacker, J. High Energy Phys. {\bf 03} (2015) 169. 

\bibitem{AFFalk} A. F. Falk, Y. Grossman, Z. Ligeti and A. A. Petrov, Phys. Rev. D {\bf 65}, 054034 (2002).

\bibitem{HYCheng} H. Y. Cheng and C. W. Chiang, Phys. Rev. D {\bf 81}, 114020 (2010).

\bibitem{MarKIII} D. Coffman {\it et al}. (Mark III Collaboration), Phys. Rev. D {\bf 45}, 2196 (1992). 

\bibitem{E691} J. C. Anjos {\it et al}. (E691 Collaboration), Phys. Rev. D {\bf 46}, 1941 (1992).

\bibitem{datasample} M. Ablikim {\it et al}. (BESIII Collaboration), Chin. Phys. C {\bf 37}, 123001 (2013).

\bibitem{datasample2} M. Ablikim {\it et al}. (BESIII Collaboration), Phys. Lett. B {\bf 753}, 629 (2016).

\bibitem{Zou} B. S. Zou and D. V. Bugg, Eur. Phys. J. A {\bf 16}, 537 (2003).

\bibitem{detector} M. Ablikim {\it et al}. (BESIII Collaboration), Nucl. Instrum. Methods Phys. Res., Sect. A {\bf 614}, 345 (2010). 

\bibitem{sim} S. Agostinelli {\it et al}. (GEANT4 Collaboration), Nucl. Instrum. Methods Phys. Res., Sect. A {\bf 506}, 250 (2003). 

\bibitem{KKMC} S. Jadach, B. F. L. Ward, and Z. Was, Phys. Rev. D {\bf 63}, 113009 (2001).

\bibitem{FSR} E. Barberio and Z. Was, Comput. Phys. Commun. {\bf 79}, 291 (1994).

\bibitem{EvtGen} D. J. Lange, Nucl. Instrum. Methods Phys. Res., Sect. A {\bf 462}, 152 (2001);\\
                 R. G. Ping, Chin. Phys. C {\bf 32}, 599 (2008).

\bibitem{LundCharm} J. C. Chen {\it et al}., Phys. Rev. D {\bf 62}, 034003 (2000).

\bibitem{KsKPi} J. Insler {\it et al}. (CLEO Collaboration), Phys. Rev. D {\bf 85}, 092016 (2012).

\bibitem{Blatt} S. U. Chung, Phys. Rev. D {\bf 48}, 1225 (1993); {\bf 57}, 431 (1998); \\
                F. von Hippel and C. Quigg, Phys. Rev. D {\bf 5}, 624 (1972). 

\bibitem{GS} G. J. Gounaris and J. J. Sakurai, Phys. Rev. Lett. {\bf 21}, 244 (1968).

\bibitem{KPiS} B. Aubert {\it et al}. ($BABAR$ Collaboration), Phys. Rev. D {\bf 78}, 034023 (2008).

\bibitem{LASS} D. Aston {\it et al}. (LASS Collaboration), Nucl. Phys. {\bf B296}, 493 (1998).

\end{thebibliography}
\end{document}